\documentclass[preprint2]{pp7}

\usepackage{color}

\newcommand{\mplanet}{$M_{\rm p}$}
\input{pp7.h}
\hypersetup{draft}
\begin{document}

\title{\textbf{\LARGE Planet-Disk Interactions and Orbital Evolution}\index{Disk-planet interactions}\index{Disks!Protoplanetary disks}\index{Giant planets!Orbital evolution}\index{Migration}\index{Planet migration}\index{Planets!migration}\index{Tidal torques}}

\author {\textbf{\large Sijme-Jan Paardekooper}}
\affil{\small\it Queen Mary, University of London, Mile End Road, London E1 4NS, United Kingdom}
\author {\textbf{\large Ruobing Dong}}
\affil{\small\it Department of Physics \& Astronomy, University of Victoria, Victoria, BC, V8P 1A1, Canada}
\author {\textbf{\large Paul Duffell}}
\affil{\small\it  Department of Physics and Astronomy, Purdue University, 525 Northwestern Avenue, West Lafayette, IN 47907-2036, USA}
\author {\textbf{\large Jeffrey Fung}}
\affil{\small\it Clemson University, 118 Kinard Laboratory, Clemson, SC 29634, USA}
\author {\textbf{\large Fr\'ed\'eric S.~Masset}}
\affil{\small\it Instituto de Ciencias F\'isicas, Universidad Nacional Autonoma de M\'exico, Av. Universidad s/n, 62210 Cuernavaca, Mor., Mexico}
\author {\textbf{\large Gordon Ogilvie}}
\affil{\small\it Department of Applied Mathematics and Theoretical Physics, University of Cambridge, Centre for Mathematical Sciences, Wilberforce Road, Cambridge CB3 0WA, UK}
\author {\textbf{\large Hidekazu Tanaka}}
\affil{\small\it Astronomical Institute, Tohoku University, Sendai, Miyagi 980-8578, Japan}

\begin{abstract}
\baselineskip = 11pt
\leftskip = 1.5cm 
\rightskip = 1.5cm
\parindent=1pc
{\small \noindent %Abstract:
Planet-disk interactions, where an embedded massive body interacts gravitationally with the protoplanetary disk it was formed in, can play an important role in reshaping both the disk and the orbit of the planet. Spiral density waves are launched into the disk by the planet, which, if they are strong enough, can lead to the formation of a gap. Both effects are observable with current instruments. The back-reaction of perturbations induced in the disk, both wave-like and non-wavelike, is a change in orbital elements of the planet. The efficiency of orbital migration is a long-standing problem in planet formation theory. We discuss recent progress in planet-disk interactions for different planet masses and disk parameters, in particular the level of turbulence, and progress in modeling observational signatures of embedded planets.      
 \\~\\~\\~}
 %leave this in to get the correct vertical space after the abstract
\end{abstract}

\section{Introduction}
The subject of planet-disk interactions has arguably entered its fifth decade \citep{1980ApJ...241..425G}. The interplay between a massive perturber and a gaseous disk, and the resulting orbital evolution of embedded planets, nevertheless remains a rich subject in which significant progress has been made since the last Protostars and Planets chapter \citep[PPVI,][]{2014prpl.conf..667B}. In addition, the subject has been propelled forward through new observations of both fully formed planets and protoplanetary disks. Below, we first provide some relevant highlights of progress in both of these observational areas since PPVI.

\subsection{Planet population}
The population of known extrasolar planets has grown dramatically since the last Protostars and Planets review \citep{2014prpl.conf..667B}. We now have more than $4,000$ confirmed planets, amongst which around our nearest neighbour star \citep{2016Natur.536..437A}, and this wealth of data reveals interesting patterns that may have bearing on planet formation and disk-planet interactions. Of particular interest are the radius valley \citep{2017AJ....154..109F}, and the eccentricity dichotomy \citep{2016PNAS..11311431X}, where single-planet systems are observed to have higher eccentricities than multiplanet systems. These structures provide valuable links between the current architecture of planetary systems and their formation \citep[e.g.][]{2017ApJ...847...29O,2018MNRAS.476..759G,2020MNRAS.498.5166P}. A more direct link to formation can be made by observing exoplanets around very young, pre-main-sequence, stars. Such young planets have been detected by the radial velocity method around weak-line T-Tauri stars V830 Tau \citep{2016Natur.534..662D}, TAP-26 \citep{2017MNRAS.467.1342Y}, and using transits for K2-33 \citep{2016Natur.534..658D}, V1298 Tau \citep{2019AJ....158...79D} and AU Mic \citep{2020Natur.582..497P}.

Other new developments include characterizing exoplanet atmospheres, again with a possible link to their formation \citep[e.g.][]{2018haex.bookE.104M}, the ability of GAIA to weigh very young planets \citep{2018NatAs...2..883S}, and the fact that the time baselines for exoplanet detections are now long enough to give information on planet occurrence beyond the snow line \citep{2021ApJS..255...14F}. This ever stronger link between observed planetary systems and their formation makes the study of disk-planet interactions, which will determine the orbital architecture of planetary systems before the protoplanetary disk disappears, ever more relevant.

\subsection{Disk observations}
Since PPVI, spatially resolved observations of protoplanetary disks have witnessed significant progress. Detailed reviews of this progress can be found in the chapter led by M. Benisty, J. Bae, and also \citet{andrews20}. Here we briefly introduce disk observations that are most relevant to detecting signatures of disk-planet interactions.

Today, high fidelity imaging observations of protoplanetary disks on the global scale with the highest angular resolutions ($\lesssim 0.1\arcsec$) and sensitivities are mainly carried out at two spectral windows. At optical to near-infrared (ONIR) wavelengths, 
a few 8-m class ground based telescopes equipped with adaptive optics (AO) and stellar halo suppression techniques have enabled us to image disks in both total and polarized intensity. A few representative instruments of this category include the Spectro-Polarimetric High-Contrast Exoplanet Research (SPHERE) on the VLT \citep[e.g.,][]{benisty15}, the Gemini Planet Imager (GPI) on Gemini \citep[e.g.,][]{follette17}, and the Subaru Coronagraphic Extreme Adaptive Optics (SCExAO) on Subaru \citep[e.g.,][]{currie19}. 
The diffraction limited angular resolution --- $\sim$40 mas, or $\lesssim6$ AU at a typical distance of 140 pc to the closest star forming regions --- can usually be achieved in the $H$-band.
At these wavelengths disks are usually extremely optically thick out to $\sim100$ AU. Such observations probe scattered starlight from the disk surface determined by the spatial distribution of small dust particles, $\sim\micron$-sized or smaller. Such particles tend to be well-mixed with the gas and have an extended vertical distribution instead of settling to the disk midplane.

The second spectral window is at millimeter (mm) to centimeter (cm) wavelengths, which has been revolutionized recently largely thanks to the Atacama Large Millimeter/submillimeter Array (ALMA)\index{Atacama Large Millimeter Array} and the upgraded Very Large Array (VLA).
Dust continuum and gas line emission can be imaged by the two interferometers. With the former, we can trace the spatial distribution of dust particles of sub-mm to cm sizes, as they dominate thermal emission at these wavelengths. The latter type of observations allows us to probe the spatial distribution, temperature, and kinematics of gas in disks.
Angular resolutions up to $\sim20$ mas, or 3 AU at a typical distance, can now be routinely achieved since the completion of the ALMA long baseline configurations \citep{brogan15}. 

\subsection{Disk properties}

Protoplanetary disks are complex entities, with different physical processes important at different locations (see chapter by Lesur et al.). Here, we start from very simple disk models, and summarise the main disk parameters that are important to understand disk-planet interactions. This sets the scene for discussions of recent progress using more realistic disk models in section \ref{sec:progress}.

Protoplanetary disks consist mostly of gas, and the relevant equations describing their dynamics are therefore given by the equations of fluid dynamics: conservation of mass, momentum and energy\index{Conservation laws}. These need to be closed by an equation of state, expressing pressure in terms of density and energy. If a barotropic equation of state is adopted, $p=p(\rho)$, then there is no need to solve the energy equation. Such approximations are still widely used, in particular the locally isothermal approximation, where the temperature is a prescribed function of radius. The simple, locally isothermal disk model can be made more realistic by including additional physics in the form of thermal effects (heating and cooling), magnetic fields (including non-ideal MHD) and solid particles in the form of dust.

Disks are assumed to be close to vertical hydrostatic equilibrium, so that the vertical component of stellar gravity is balanced by a pressure gradient. If the disk is vertically isothermal, this leads to a Gaussian density profile with a pressure scale height $H$ that is determined by the temperature (or sound speed, $c_s$):
\begin{align}
H = c_s/\Omega_{\rm K},
\end{align}
where $\Omega_{\rm K}$ is the local Keplerian angular velocity. Protoplanetary disks are usually thin, so that $H/r \equiv h \ll 1$. This allows in some cases for a simplification through vertical averaging, after which we can effectively work in two dimensions and with a surface density instead of volume density.  

Disk masses are typically much smaller than the stellar mass, and the Toomre $Q$ parameter is much larger than unity for large parts of the lifetime of the disk. This means that self-gravity can be safely neglected. This is not true for younger disks, which are thought to go through a phase where they are self-gravitating and potentially form giant planets by gravitational instabilities \citep[e.g.][]{1978M&P....18....5C, 1997Sci...276.1836B}. It should also be noted that in some important cases self-gravity can have important effects when $Q\gg 1$ but $Qh \sim 1$. This includes the Type I migration rate \citep{2008ApJ...678..483B}, global edge modes induced by planetary gaps \citep{2011MNRAS.415.1445L}, and vortices in low-viscosity disks \citep{zhu16lopsided, lega21}.

A small percentage of the mass of the disk is in solid form, which is usually called \emph{dust}. The canonical value for the dust-to-gas ratio is $1/100$ \citep{2011piim.book.....D}, which might indicate that dust is dynamically unimportant. However, dust tends to cluster in pressure maxima, so that the local dust-to-gas ratio may be much higher than $1/100$. This is where dust can in principle play a role in disk-planet interactions.\index{Dust!Gas drag}

Finally, angular momentum transport\index{Angular momentum transport}\index{Disks!Angular momentum transport} in disks is usually modelled by including a turbulent viscosity, usually through the $\alpha$-prescription \citep{1973A&A....24..337S}:
\begin{align}
    \nu = \alpha c_s H,
\end{align}
where $\nu$ is the kinematic viscosity and $\alpha$ is a dimensionless constant. The underlying idea is that while molecular viscosity is many orders of magnitude too small to play a role, the disk is likely to be turbulent, with a maximum eddy size of $\sim H$ (to fit into the disk) and maximum turbulent velocities of $\sim c_s$ (otherwise strong damping by shocks). One then expects $\alpha < 1$. While there are obvious limitations to modelling turbulence this way, the level of viscosity serves as a useful boundary for various regions of disk-planet interactions (see section \ref{sec:regimes}).

\subsection{Outline}
The rest of the chapter is organised as follows. We begin in section \ref{sec:regimes} by providing a basic overview of the various regimes of disk-planet interactions. We then go on to discuss areas of progress since the last Protostars and Planets conference in section \ref{sec:progress}. We discuss the tools used to make progress in section \ref{sec:tools}. In section    \ref{sec:obs} we discuss possible observational signatures of disk-planet interactions, and conclude in section \ref{sec:con}.

\begin{figure}[h]
 \epsscale{1.0}
 \plotone{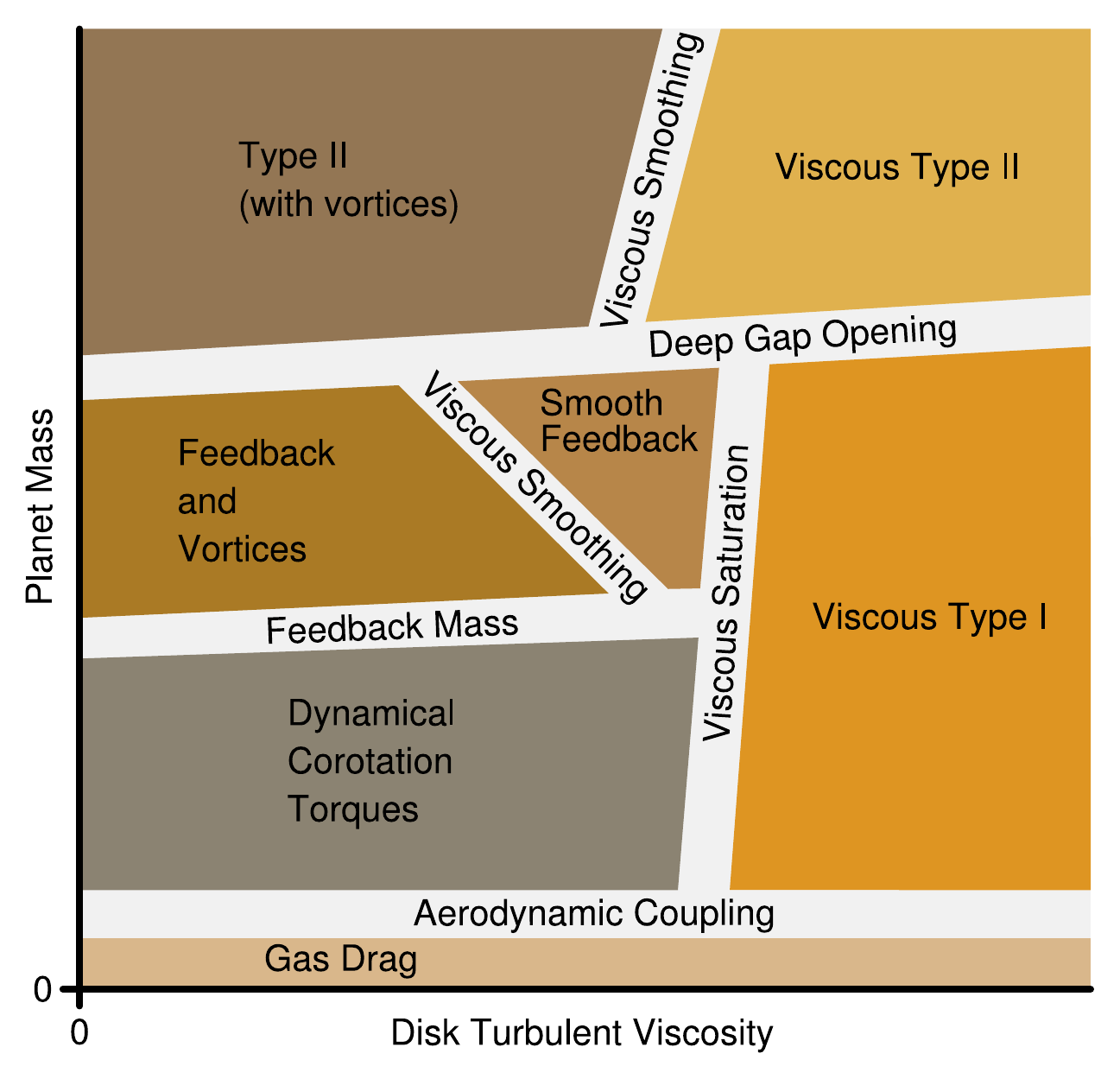}
 \caption{\small Organization of basic planet migration regimes from \citet{2019MNRAS.484..728M}.}
 \label{fig:migration_org_chart}
\end{figure}

\section{Overview of Planet-Disk Interaction Regimes}
\label{sec:regimes}

The interaction of a body orbiting a young star in the protoplanetary disk has a different character as a function of the parameters of the solid body, the orbital trajectory, and the disk. To establish a rough orientation in the dominant physical regimes and the scales between them we use the map in Figure~\ref{fig:migration_org_chart}.

We first summarize the basic flow structures in the disk resulting from an embedded planet. We then go on to describe the various regions of Figure~\ref{fig:migration_org_chart} that were well known at the time of the last Protostars and Planets review.

\subsection{Flow structure around an embedded object\index{Disks!Spiral waves}}

\begin{figure}[ht]
 \epsscale{1.0}
 \plotone{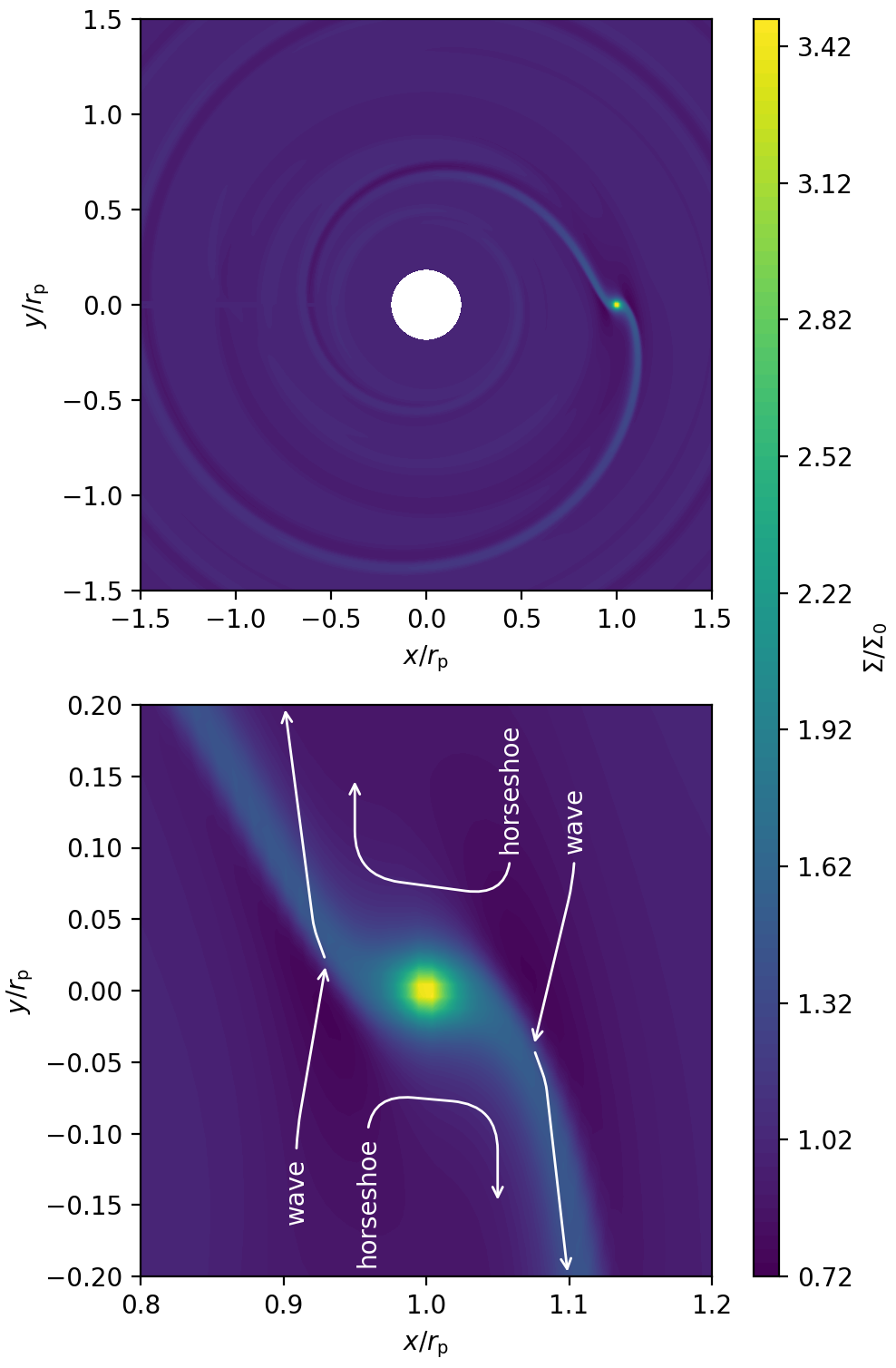}
 \caption{\small Surface density perturbation around a Neptune-mass planet ($q=10^{-4}$) on a circular orbit. Top panel: global overview of the disk, showing the prominent one-armed spirals. Bottom panel: zoom-in on the planet, with schematic streamlines (in the frame rotating with the planet) superimposed, indicating horseshoe trajectories as well as trajectories responsible for the wave torque.}
 \label{fig:flow}
\end{figure}

The response of a gaseous disk to a massive embedded object can be divided up into a wavelike and a non-wavelike response. The wavelike response can be thought of to be generated at Lindblad resonances \citep{1979ApJ...233..857G}\index{Disks!Resonances}\index {Lindblad resonances}\index{Planet formation!Resonances} and results in a one-armed spiral due to constructive interference of Fourier modes \citep{2002MNRAS.330..950O}. An example of a Neptune-sized planet (mass ratio $q \equiv M_{\rm p}/M_*=10^{-4}$) embedded in a disk with $H/r=0.05$ is shown in Figure \ref{fig:flow}, showing the prominent one-armed spiral density waves in the top panel.
Additional spirals may emerge at larger distances from the planet, depending on the disk and planet properties \citep{bae18}. Such spiral patterns have been observed in a number of disks in scattered light as well as dust and gas emission (see section \ref{sec:obs}).
%{\color{red} Ruobing: perhaps add a few sentences if necessary for your observations discussion?}

Close to the orbit of the planet, there is a wave-free zone if the disk is not self-gravitating and only weakly magnetised  \citep[e.g.][]{1979ApJ...233..857G, 2003MNRAS.341.1157T,2013MNRAS.430.1764G,2015ApJ...802...54U}. This is where the fluid executes horseshoe turns in the frame corotating with the planet \citep[e.g.][]{1981Icar...48....1D}. In a gas disk, the width of the horseshoe region depends both on the mass of the planet and the temperature of the disk \citep{2009MNRAS.394.2297P}. For low-mass planets, the half-width of the horseshoe region $x_\mathrm{s}/r_\mathrm{p} \sim \sqrt{q/h}$, while for high-mass planets the gas-free result is obtained, $x_\mathrm{s}/r_\mathrm{p} \sim q^{1/3}$. The connection between the two mass regimes is such that for intermediate-mass planets the horseshoe region is wider than would be expected based on the low-mass case \citep{2006ApJ...652..730M}.  \citet{2017MNRAS.471.4917J} provide an expression of the width of the horseshoe region suitable in this transitional regime. In the bottom panel of Figure \ref{fig:flow} a close-up of the planet is shown, with schematic streamlines for both wavelike perturbations and horseshoe turns.

\subsection{Gas Drag / Type 0\index{Dust!Gas drag}} 
Bodies smaller than planetesimals of $\sim 10$~km drift due to gas drag (see chapter by Drazkowska et al.). In this case, gravitational interactions do not play a role. This regime is indicated at the bottom of Figure~\ref{fig:migration_org_chart}. While for larger bodies, gravitational interactions usually dominate over aerodynamics, in specific cases planetary bodies may undergo Type 0 migration \citep{2009ApJ...702L.182A, 2009A&A...506L...9P, 2021arXiv210703413B}. This kind of frictional interaction is also relevant for highly eccentric or inclined orbits \citep{2012MNRAS.422.3611R}.

\subsection{Viscous Type I\index{Planet migration!Type I}\index{Terrestrial planets!Formation}} 
Moving up in planet mass above the {\sl aerodynamic coupling} transition in Figure~\ref{fig:migration_org_chart}, gravitational interaction with the disk becomes the dominant source of torques. This is dominated by the wave torque \citep{1980ApJ...241..425G, 1993ApJ...419..155A} and the corotation torque or horseshoe drag \citep{1991LPI....22.1463W}. Towards larger values of turbulent viscosity, this is the realm of classic, viscous Type I migration \citep{1997Icar..126..261W, 2002ApJ...565.1257T}. It is characterised by a corotation torque that is kept unsaturated by turbulent viscosity \citep{2001ApJ...558..453M,2004ApJ...615.1000M,2007LPI....38.2289W, 2010ApJ...723.1393M, 2011MNRAS.410..293P}. This means that while the reservoir of angular momentum that coorbital material can give to the planet is finite, viscosity is strong enough to replenish the angular momentum before it runs out. Towards lower values for the viscosity, coorbital angular momentum is not replenished fast enough, which leads to a reduced, saturated corotation torque. Towards higher planet masses, torques on the disk become strong enough to modify the density structure of the disk, leading to the opening of a deep annular gap, which is the realm of Type II migration. 

A significant amount of effort has been made over the years to better understand Type I migration and to come up with parametrizations that can be used for example in N-body simulations and planet population synthesis models. Early examples include \cite{1993Icar..102..150K}, who tackled the two-dimensional isothermal case, and \cite{2002ApJ...565.1257T}, who provided torque prescriptions for three-dimensional, isothermal disks. These have the following form for the torque on the planet:
\begin{align}
    \Gamma = C \frac{q^2}{h^2} \Sigma_{\mathrm{p}} r_{\mathrm{p}}^4 \Omega_{\mathrm{p}}^2,
    \label{eq:torque_on_planet}
\end{align}
where $q$ is the mass ratio planet/star, $h = H/r$ is the aspect ratio of the disk, $\Sigma_{\mathrm{p}}$ is the surface density, $r_{\mathrm{p}}$ the orbital radius and $\Omega_{\mathrm{p}}$ the orbital frequency. All spatially varying quantities are evaluated at the location of the planet, as indicated with a subscript $p$. The dimensionless constant $C$ depends on the background density and temperature profiles and is typically of order unity and negative, implying inward migration. It should be noted that unless the viscosity is strong enough to directly affect the horseshoe turns, the corotation torque is given by (nonlinear) horseshoe drag rather than the linear corotation torque \citep{2009MNRAS.394.2283P,2009ApJ...703..845C}.

When the assumption of an isothermal disk is relaxed, the corotation torque starts to play a very prominent role \citep{2006A&A...459L..17P,2008ApJ...672.1054B,2008A&A...485..877P,2008A&A...487L...9K}. Based on the local gradient of entropy in the disk, it can even drive outward migration if the cooling of the disk is sufficiently inefficient for the disk to behave almost adiabatically \citep[e.g.][]{2009ApJ...703..857M,2010MNRAS.401.1950P}. The constant $C$ in equation (\ref{eq:torque_on_planet}) then depends on the local gradients of density and temperature, the level of turbulence, which governs the saturation of the corotation torque, and the cooling time of the disk. Parametrizations of the torque on the planet in this regime were presented in \citet{2010ApJ...723.1393M} and \citet{2011MNRAS.410..293P}, and subsequently by \citet{2017MNRAS.471.4917J} who provided torque formulae adapted to three-dimensional disks.

\subsection{Viscous Type II\index{Planet migration!Type II}} 
\label{sec:typeII}

%Conceptually, with a planet massive enough to effectively empty a gap ({\sl Deep Gap Opening}), the planet's migration becomes locked to the background viscous evolution of the disk \citep{1986ApJ...309..846L}. Significant effort has been expended on refining this scenario, and it will be discussed at length in Section~\ref{sec:typeII}.

%As planet masses increase, dissipation of the spiral wake, generated by waves launched from Lindblad resonances, is able to modify the disk surface density profile. The mass scale at which this occurs depends at least on the disk viscosity parameter. In this section we discuss gap opening mass scales in viscous conditions. This may be discussed in detail in a separate chapter on rings in protoplanetary disks, but for our purposes the criteria from  \citep{2015ApJ...806L..15K} is sufficient for viscous disks.

%In the inviscid limit, a mass scale can be derived by finding the mass where a planet driven by the classical Type I torque opens a gap faster than it migrates past \citep{1989ApJ...347..490W,2002ApJ...572..566R}, which is shown in Figure~\ref{fig:migration_org_chart} as the Feedback Mass.

When the planet becomes sufficiently massive, the spiral waves emanating from the planet deposit a significant amount of torque in the disk (typically via shocks), carving out a  low-density annulus (i.e., a planetary gap) along the planetary orbit. The classical model of Type II migration assumed the density inside the gap to be low enough such that Type I torques play no role, and no gas flows crossing the gap. Under these assumptions, the planet motion would be locked to the same drift velocity as the background viscous accretion\index{Accretion} of the disk \citep[e.g.][]{1986ApJ...309..846L}. In other words, the planet is locked inside its own gap, which slowly accretes onto the star.

However, many hydrodynamical simulations have observed considerable gas flows crossing the planetary gaps since the end of the last century \citep[e.g.][]{1996ApJ...467L..77A,1999ApJ...514..344B,1999MNRAS.303..696K,1999ApJ...526.1001L,2001MNRAS.320L..55M}.  Gas was found to easily drift through the gap region on horseshoe orbits.  Recent simulations on Type II migration also found that a massive planet with 
a gap can migrate independently of the disk accretion, owing to the gap-crossing flow 
\citep[e.g.][]{2014ApJ...792L..10D, 2015A&A...574A..52D, 2017A&A...598A..80D}.
A planet migrates together with the gap because the planet keeps digging a gap at its position.
Based on these hydrodynamical simulations, a new simple model of Type II migration was proposed by \citet{2018ApJ...861..140K}, to replace the classical model.  This new migration model is dependent upon the accurate gap structure revealed by recent hydrodynamical simulations 
\citep[e.g.][]{2013ApJ...769...41D,2014ApJ...782...88F,2015MNRAS.448..994K}.
Such recent progress on the nature of gap structure and Type II migration will be discussed at length in Section~\ref{sec:nature_of_typeII}.

There is some ambiguity in the semantic question of what ``Type II" actually means.  In some cases, ``Type II migration" has been used to mean that the planet migrates in lockstep with the viscous drift rate of the disk \citep[indeed, that is how the regime ``Type II" was defined when it was first coined, see][]{1997Icar..126..261W}.  Under this definition, ``Type II" migration would never happen, according to our present understanding.  However, in this review we define ``Type II" to simply mean the regime in which the planet opens a significant gap, enough to change the planet's migration rate.

At what planet mass does this transition occur?  The criterion proposed by \citet{2006Icar..181..587C} had been widely used for many years, which is a combination of the thermal condition and the viscous condition.  However, this criterion is no longer considered accurate, especially for low-viscosity disks.
The gap-opening criterion has since been modified by recent hydrodynamical simulations \citep[e.g.][]{2013ApJ...769...41D}
and by traditional analytical considerations \citep[e.g.][]{2013ApJ...768..143Z,2014ApJ...782...88F,2015MNRAS.448..994K}.
The presently-understood gap opening condition is the same as the traditional viscous criterion 
\citep[e.g.][]{1986ApJ...309..846L,2002ApJ...572..566R}
derived from the balance between the angular momentum flux of three-dimensional spiral waves excited by the planet \citep[e.g.][]{2002ApJ...565.1257T} and the angular momentum flux of the viscous accretion disk.
The classical thermal condition (i.e. $R_{\mathrm{Hill}} \gtrsim H$) is found to be unnecessary, 
as previously suggested by \citet{2002ApJ...572..566R}.  Thus, the gap opening condition is expressed as
\citep[][]{2013ApJ...769...41D,2015MNRAS.448..994K}
\begin{align}
q \gtrsim 5 h^{3/2} \left( \frac{\nu}{r_{\mathrm{p}}^2 \Omega_{\mathrm{p}}}\right)^{1/2} = 5 h^{5/2} \alpha^{1/2}.
\label{eq:gap_opening_condition}
\end{align}
This gap opening condition indicates that even a Neptune-mass planet will open a gap in a disk with a low-viscosity 
of $\alpha \lesssim 10^{-4}$ and a disk aspect ratio of $h=0.05$. The gap opening by such a low-mass planet was also 
directly shown by numerical simulations by \citet{2013ApJ...769...41D}, 
although this gap opening had been excluded by the classical thermal condition.

Even when the gap opening condition is satisfied, the gap does not become empty immediately but the gap surface density 
decreases with a power-law of $q^{-2}$ \citep[e.g.][]{2013ApJ...769...41D}.  As the gap becomes sufficiently deep, the gap depth becomes a steeper function of $q$ \citep[]{2014ApJ...782...88F, 2020ApJ...901...25D}.  The low-density gas inside the gap causes gap-crossing flows and  accretion flows onto the planet 
even in a case of a Jupiter-mass planet. 
It also enables the planet to migrate independently of the disk accretion.  

For a viscous disk, the gap-opening criterion can be considered as a competition between timescales; i.e. can the planetary torques open the gap faster than viscosity can refill the gap?  In the inviscid case, a different condition to equation (\ref{eq:gap_opening_condition}) should apply.  In the absence of viscosity, gap opening is a competition between planetary torques and migration; i.e. can the planet open the gap on a timescale shorter than its migration timescale?  This criterion is known as the ``inertial limit" \citep{1989ApJ...347..490W,2002ApJ...572..566R,2015ApJ...802...56M}, which is shown in Figure~\ref{fig:migration_org_chart} as the Feedback Mass.  In principle, other timescales could be compared with the gap-opening timescale, e.g. the lifetime of the disk, but the inertial limit is most likely the relevant criterion in the inviscid case.

\subsection{Type III\index{Planet migration!Type III}}

Viscous Type I migration is driven by asymmetries between the inner and outer spiral wave (the wave torque) and asymmetries between the two horseshoe legs due to radial gradients in density and temperature. Torque prescriptions are usually derived by keeping the planet on a fixed orbit and measuring the steady state torque on the planet \citep{2010ApJ...723.1393M, 2011MNRAS.410..293P}. If the planet is migrating, this potentially creates another source of asymmetry between the two wakes and in particular the two horseshoe legs.  

\begin{figure}[ht]
 \epsscale{0.5}
 \plotone{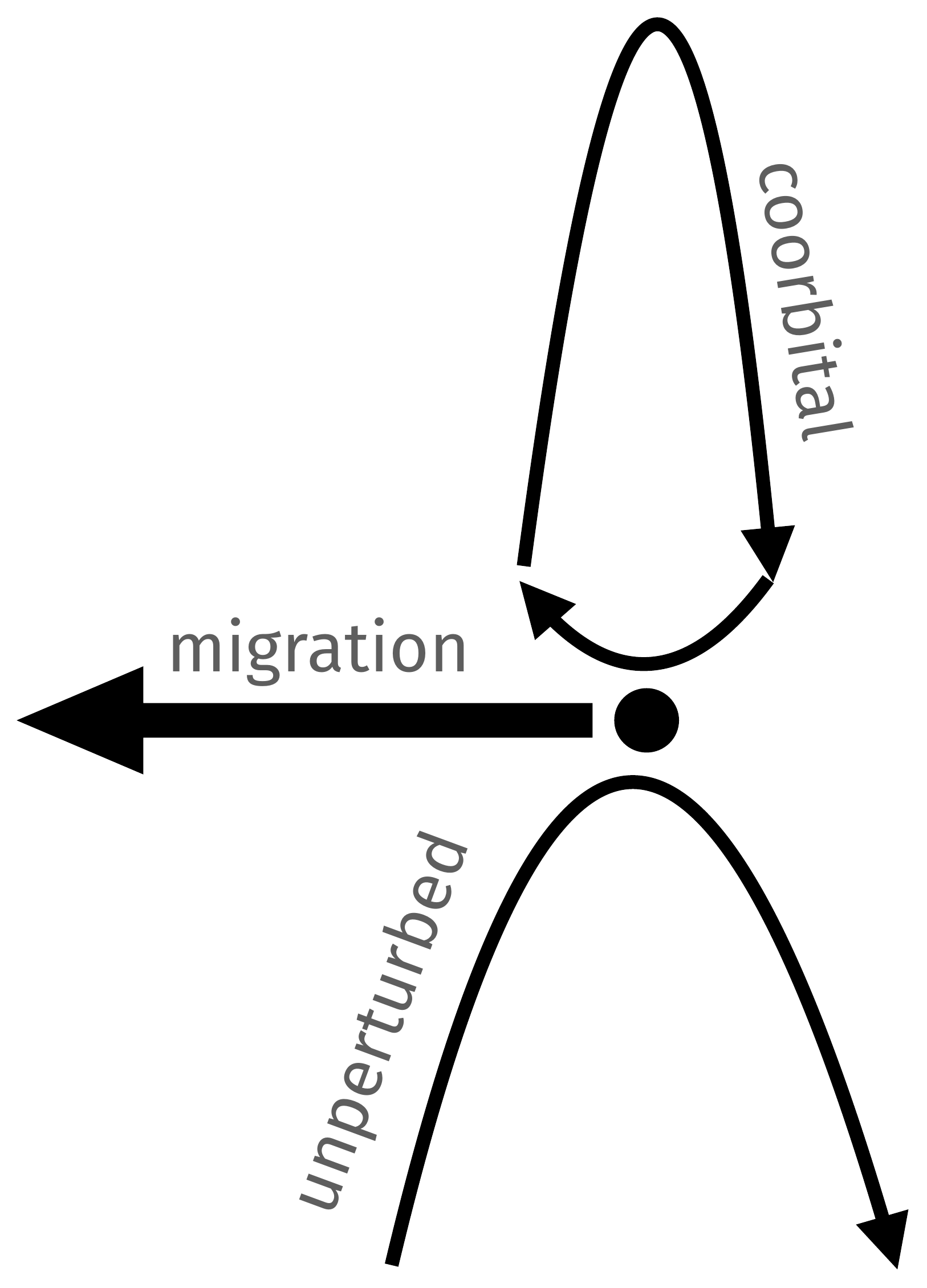}
 \caption{\small Schematic flow structure of the coorbital region in the comoving frame of a migrating planet. Orbital radius varies horizontally, while the direction of orbital motion is vertical. The planet is migrating towards the left of the figure.}
 \label{fig:typeiii}
\end{figure}

The basic picture is illustrated schematically in Figure \ref{fig:typeiii}. While in the non-migrating case, both horseshoe legs would be symmetric, if the planet is migrating towards the left of the figure, all material not interacting with the planet will drift towards the right in the comoving frame. This leads to the structure depicted in Figure \ref{fig:typeiii}. On the bottom leg, unperturbed disk material streams past the planet, executing only a single horseshoe turn. The top leg consists of coorbital material that migrates with the planet.

This setup was first analysed in \citet{2003ApJ...588..494M}. If the planet has partially depleted its coorbital region, this will lead to a different torque from both legs even in the absence of global gradients. Moreover, when migration is slow compared to the horseshoe libration time, the resulting torque is proportional to the migration rate, opening up the possibility of a runaway process \citep{2003ApJ...588..494M, 2004pfte.confE...2A}. This was studied further in \cite{2008MNRAS.386..164P,2008MNRAS.386..179P,2008MNRAS.387.1063P}. For a more in-depth discussion see \cite{2007prpl.conf..655P}.

%
%Other discussions will be deferred to the following section on recent progress.

%where to cite  Application of Gas Dynamical Friction for Planetesimals. I. Evolution of Single Planetesimals 2015ApJ...811...54G if at all?

\section{Areas of Recent Progress}
\label{sec:progress}
In this section, we review areas where there has been significant progress and interest, particularly since PPVI \citep{2014prpl.conf..667B}.

\subsection{Discovery of Thermal Torques\index{Disks!Thermodynamics}\index{Planet migration!Type I}\index{Terrestrial planets!formation}}
\label{sec:thermal}
Most effects on the planetary torque arising from thermal diffusion within the gaseous disk, and those due to the release of energy into the surrounding gas by an accreting, low-mass planet, were not taken into consideration in the work reported in PPVI. Since then, it has been discovered \citep{2014MNRAS.440..683L} that the diffusion of heat in the vicinity of a low-mass planet can have a large impact on the disk torque. The gas that tends to accumulate in the vicinity of the planet is compressionally heated. The heat thus produced tends to diffuse outwards, and the planet's neighbourhood is colder (hence denser, to maintain the pressure balance) than it would be if the gas behaved adiabatically. The difference in gas density with respect to the adiabatic case takes therefore the form of a dense peak centred on the planet. At small distance from the latter, diffusion occurs on a timescale much shorter than that of advection, and this peak has a central symmetry with a density that decays as $r^{-1}$, $r$ being the distance to the planet. At further distances, the diffusion timescale increases and becomes larger than the shear timescale. Advection then takes over and distorts the perturbed region: for a planet on a circular, non-inclined orbit, two lobes appear downstream of the planet (therefore one behind the planet in the outer disk and one leading the planet in the inner disk). As the disk is generally slightly sub-Keplerian, the planet is outside its corotation, so that the outer lobe is more pronounced and dominates the force exerted on the planet. Since this dense lobe is trailing the planet, its contribution to the planetary torque is negative: in a disk with thermal diffusion, inward migration is faster than it would be in a similar, adiabatic disk. Owing to the shape of the density difference with respect to the adiabatic case, \citet{2014MNRAS.440..683L} have called this effect the \emph{cold finger effect}. The new torque contribution arising from the lobes has been called \emph{thermal torque} \citep{2017MNRAS.472.4204M}. In the case studied by \citet{2014MNRAS.440..683L}, where the planet does not release energy into the ambient disk, it is more specifically called the \emph{cold thermal torque}.  The typical size $\lambda$ of the lobes can be obtained by equating the diffusion timescale $\lambda^2/\chi$ (where $\chi$ is the thermal diffusivity) and the shear timescale $\Omega^{-1}$. One obtains:
\begin{equation}
  \label{eq:sizethermallobes}
  \lambda\sim\sqrt{\chi/\Omega}.
\end{equation}
In planet-forming regions of protoplanetary disks, radiative diffusion is by far the main source of heat diffusion, and the thermal diffusivity takes the form:
\begin{equation}
  \label{eq:chiexpr}
  \chi = \frac{16(\gamma-1)\sigma T^3}{3\rho^2({\cal R}/\mu)\kappa},
\end{equation}
where $\gamma$ is the adiabatic index, $\sigma$ is Stefan's constant, $T$ the midplane temperature, $\rho$ the midplane density, ${\cal R}$ the ideal gas constant, $\mu$ the mean molar mass and $\kappa$ Rosseland's mean opacity \citep{2011MNRAS.410..293P}. Eq.~(\ref{eq:chiexpr}) has been checked by numerical experiments \citep{2017MNRAS.471.4917J} and found to yield values of the thermal diffusivity within $20$ to $30$~\% of that measured in runs involving the radial thermal diffusion of hot, radially narrow regions. Evaluating $\lambda$ in the planet forming regions (within $10$~au) of protoplanetary disks shows that it is usually a tiny fraction ($\sim O(10^{-1})$) of the pressure length scale $H$. Obtaining converged results for thermal torques in numerical simulations is therefore quite challenging and requires very high resolutions \citep{2021MNRAS.501...24C}, which explains why thermal torques have been unnoticed for so long in studies of planet-disk interactions. Despite the minute size of the region involved in thermal torques, they can have a large impact on the net torque. In particular, for very low-mass planets (typically below one Earth mass), they can be larger than the differential Lindblad torque by an order of magnitude  \citep{2017MNRAS.472.4204M}.
Note that so far there has not been any study of thermal torques in turbulent disks. Thermal torques are typically established over one orbital period \citep{2021MNRAS.501...24C}. Should the turnover time of turbulent eddies with size $\sim\lambda$ be shorter than this timescale, one could expect a sizable impact on the value of thermal torques. This question requires further study.

\begin{figure*}[h]
 \epsscale{2.0}
 \plotone{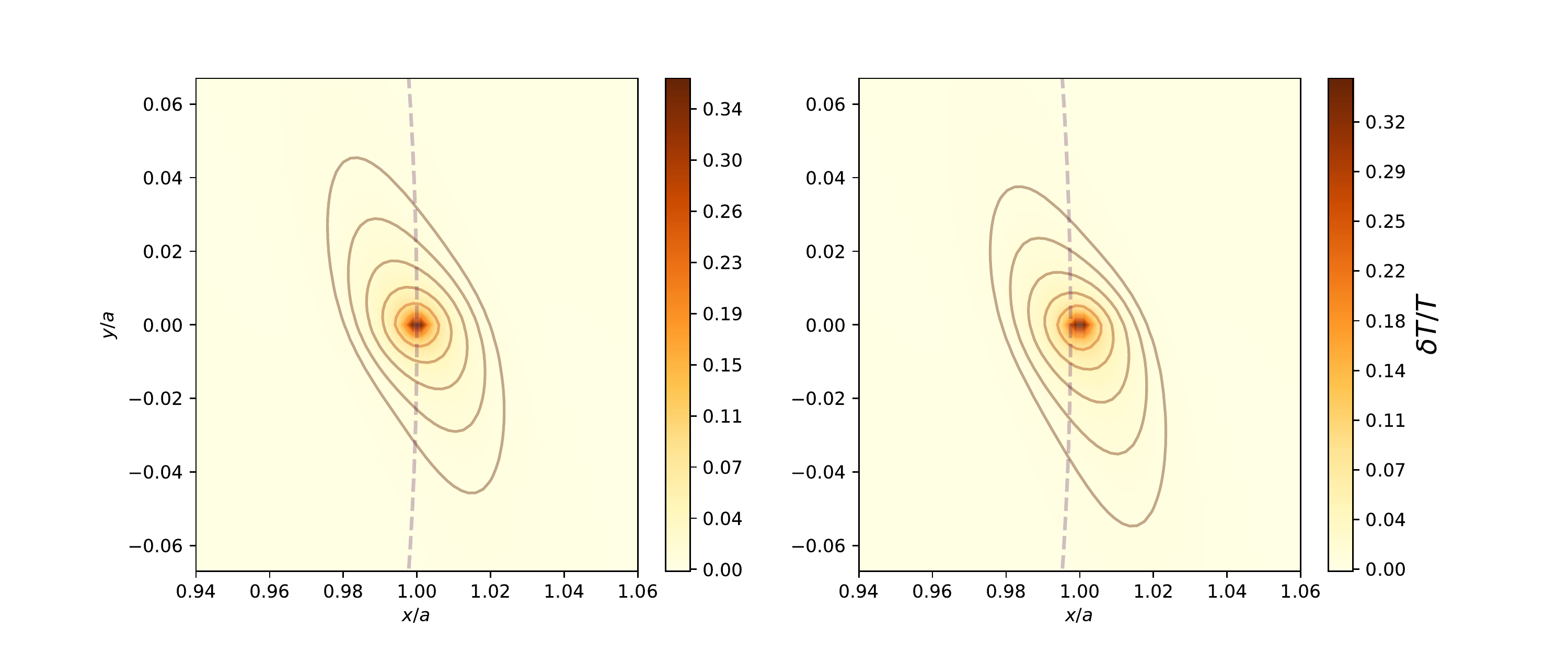}
 \caption{\small Relative perturbation  of temperature arising from heat release by a low-mass planet, at the midplane of the disk. The central star is located at $(x,y)=(0,0)$. The left plot shows the case of a disk without radial pressure gradient. The planet is then located exactly at its corotation radius (materialized by the dotted line), and the two-lobe pattern is essentially symmetric. The right plot shows the more realistic case of a slightly sub-Keplerian disk. The planet is then located outside of its corotation radius, and the two-lobe pattern is no longer symmetric. Successive isocontours differ by a factor of $2$, while the outermost contour corresponds to a value of $\delta T/T$ that is $1/64^{th}$ of the peak value obtained in these calculations. These results have been obtained using pairs of three-dimensional numerical simulations with identical parameters, one in which the planet has a luminosity $L$ and another one in which it is non-luminous. Here, the pressure length scale is $H=0.05a$, the characteristic size $\lambda$ is $\sim 0.14H$ (Eq.~\ref{eq:sizethermallobes}) and the luminosity is $L\sim 2.3L_c$, where $L_c$ is given by Eq.~\eqref{eq:lumcrit}.}
 \label{fig:TTtempfields}
\end{figure*}

\begin{figure}[h]
 \epsscale{1.0}
 \plotone{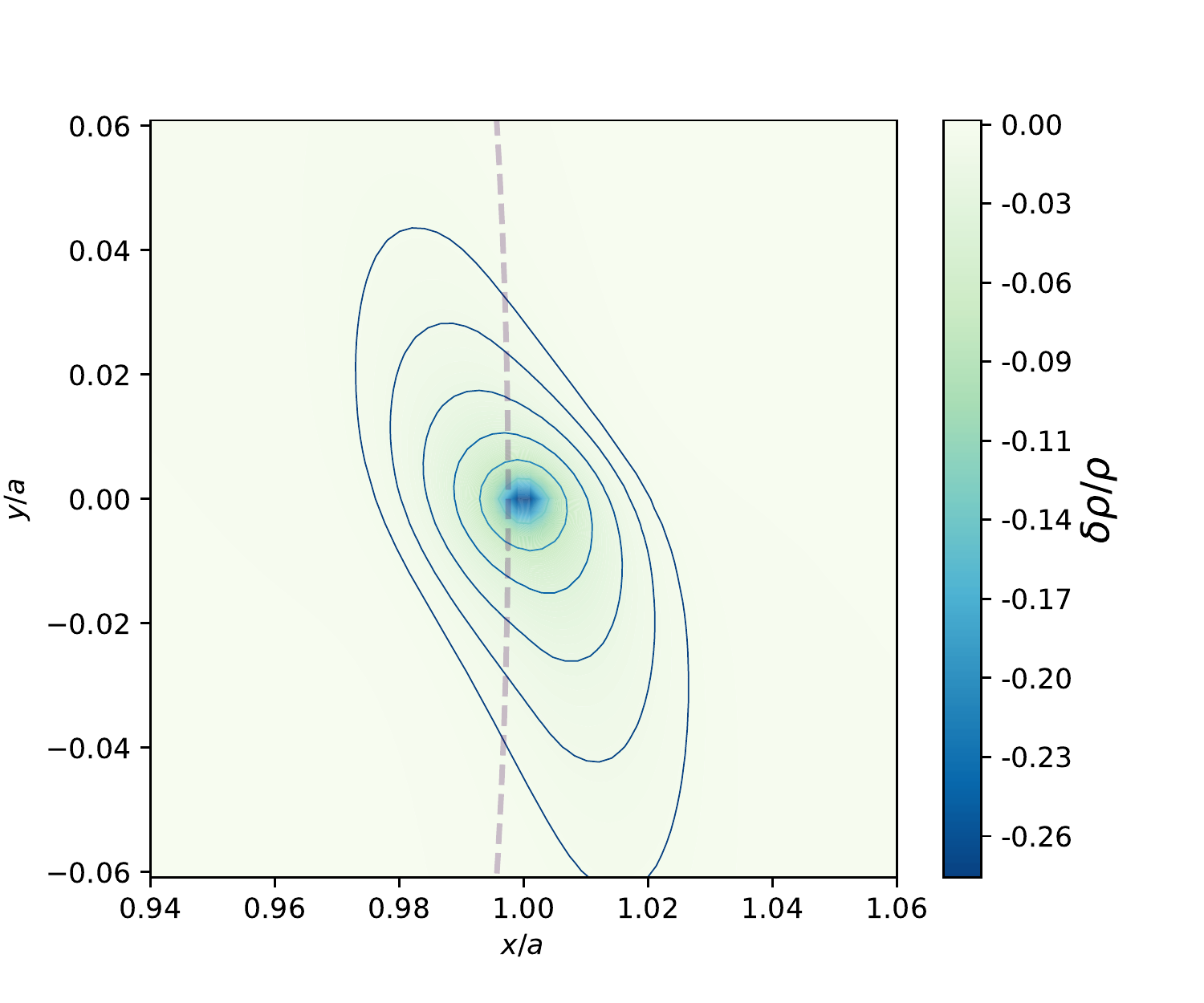}
 \caption{\small Relative perturbation of the density field arising from heat release at the midplane of the disk corresponding to the right plot of Figure~\ref{fig:TTtempfields}. The isocontours have broadly the same shape as in that figure. The sign of the perturbation is the negative of that of the temperature: the lobes are under-dense.}
 \label{fig:TTrhofield}
\end{figure}

A faster inward migration when thermal diffusion is accounted for is not the end of the story. Low-mass planets may accrete pebbles or planetesimals. When that happens, they heat the surrounding gas and a process similar to that outlined above takes place, except for a change of sign: with respect to the non-accreting case, a hot, underdense region develops, with a two-lobe shape \citep{2015Natur.520...63B}, as illustrated in Figs.~\ref{fig:TTtempfields} and~\ref{fig:TTrhofield}. In the usual case of a sub-Keplerian disk, the outer lobe dominates the net contribution to the torque. This contribution, dubbed \emph{heating} torque, is positive. If the planet is sufficiently luminous, it can reverse the torque on a low-mass planet, driving it outwards. Using linear perturbation theory, one can obtain the expression of the thermal torque, when the distance of the planet to its corotation $x_\mathrm{p}$ is much smaller than the size of the thermal lobes \citep{2017MNRAS.472.4204M}:
\begin{equation}
  \label{eq:2}
  \Gamma_\mathrm{thermal}=1.61\frac{\gamma-1}{\gamma}\frac{x_\mathrm{p}}{\lambda}\left(\frac{L}{L_c}-1\right)\Sigma_\mathrm{p}r_\mathrm{p}^4\Omega_\mathrm{p}^2q^2h^{-3}, 
\end{equation}
where $L$ is the planet's luminosity and $L_c$ the watershed luminosity at which the net thermal torque changes sign: 
\begin{equation}
  \label{eq:lumcrit}
  L_c=\frac{4\pi GM_\mathrm{p}\chi\rho}{\gamma}.
\end{equation}
When $x_\mathrm{p}$ becomes comparable to $\lambda$, the thermal torque no longer scales with $x_\mathrm{p}$ and tends to saturate to a value of order $\sqrt{2\pi}(\gamma-1)/\gamma (L/L_c-1) \Sigma_\mathrm{p}r_\mathrm{p}^4\Omega_\mathrm{p}^2q^2h^{-3}$ \citep{2021MNRAS.501...24C}. \citet{2020ApJ...902...50H} have used high-resolution, three-dimensional calculations in the shearing sheet to assess the validity of linear theory. They found an agreement at better than the $10$~\% level in the limit of low luminosity and low thermal diffusivity, and departures from linear theory at high luminosity or high thermal diffusivity.

Thermal effects also have an impact on the time evolution of eccentricity and inclination. These effects are described in section~\ref{sec:eccentr-incl-evol}.

Thermal forces are the dominant contribution to the force exerted by the disk onto the planet up to a few Earth masses. As the mass increases past this value, their relative contribution to the net force decays slowly, and becomes negligible above $10-20$ Earth masses. The decay of thermal forces has been studied in detail by \citet{2020MNRAS.495.2063V}, in a different context: that of a point-like mass that travels through a uniform three-dimensional medium with thermal diffusivity $\chi$, and that releases energy at a constant rate $L$. This setup corresponds to that of dynamical friction in a gas studied by \citet{1999ApJ...513..252O}, with the sole modifications of the inclusion of thermal diffusion in the gas, and of the energy released by the travelling mass. Owing to the existence of an axis of symmetry, this setup can be numerically tackled with two-dimensional meshes, and allows for a much more extensive study of the parameter space, and a higher resolution, than the more complex situation of a planet embedded in a stratified gaseous disk with a sheared Keplerian flow.  \citet{2020MNRAS.495.2063V} find that the thermal force exerted on the travelling mass is in good agreement with that predicted by linear theory \citep{2017MNRAS.465.3175M,2019MNRAS.483.4383V} for masses up to a few
\begin{equation}
  \label{eq:masscrit}
  M_c=\frac{\chi c_s}{G},
\end{equation}
and is $O(M_c/M)$ times smaller at larger mass.
The value of the critical mass $M_c$ depends on the location in the disk. It typically ranges from a fraction of an Earth mass just beyond the snow line to several Earth masses at $10$~au. Inside the snow line, by virtue of Eq.~(\ref{eq:chiexpr}), the critical mass can be of several Earth masses, owing to the drop of opacity.
Although no systematic study of the decay of thermal forces at larger planetary mass has been undertaken for planets embedded in protoplanetary disks, the findings of \citet{2021MNRAS.501...24C} for a planet on a circular orbit are compatible with a decay law similar to that found for dynamical friction. Similarly, the decay law for planets on eccentric or inclined orbits should closely match that found in studies of dynamical friction, since for eccentricities or inclinations larger than $\lambda/H$, the thermal response reduces essentially to a hot or cold trail, depending on the value of the luminosity \citep{2017A&A...606A.114C}. The results of \citet{2017MNRAS.469..206E}, albeit obtained at low resolution, seem to confirm this expectation.

An additional complication for planets with masses in excess of $M_c$ is that the width of their horseshoe region is comparable to or larger than the size $\lambda$ of the thermal disturbance \citep{2021MNRAS.501...24C}. The thermal lobes are therefore distorted with respect to the shape they have at lower mass. The impact of this effect on the magnitude of the thermal torque is yet to be assessed. A more extreme example of the consequences of a complex flow in the planet's vicinity has been found by \citet{2019A&A...626A.109C}, who performed numerical simulations at high resolution with a non-constant opacity, and found that the heating torque is then strongly time variable, inducing episodes of inward as well as outward migration. This arises from the advection of heat in the planet's immediate vicinity, on tightly wound streamlines. This process induces a highly time-variable configuration of the thermal lobes.

 \subsection{A Generalization of Dynamical Corotation Torques\index{Planet migration!Type I}}
 \label{sec:dynamical}
 The `Viscous Saturation' vertical division in Figure~\ref{fig:migration_org_chart} denotes where viscous diffusion of angular momentum in the disk can maintain the background gradient of disk properties across the corotation region. In the absence of viscous diffusion, phase mixing of material in the corotation region removes the background gradients that would otherwise drive the corotation torque. In this case, the contrast and asymmetry of the disk material trapped on librating orbits determines the corotation torque.  This is particularly applicable to low-turbulence wind-driven disks \citep[e.g.][]{2014A&A...566A..56L}, and has received much attention in recent years. For low-mass planets, this is the realm of dynamical corotation torques (see Figure \ref{fig:migration_org_chart}). 
 
 \begin{figure*}[ht]
 \epsscale{2.0}
 \plotone{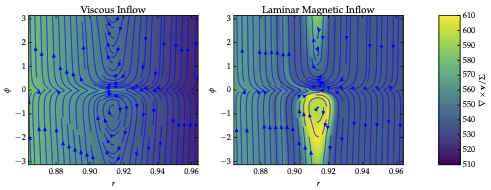}
 \caption{\small Comparison of the coorbital region for two planets at the same orbital location. Left panel: Viscously driven inflow case, the vortensity gradient across the corotation region is maintained close to the background disk value by the viscosity, resulting in an unsaturated corotation torque. Right panel: Laminar magnetic stress driven inflow case, where the vortensity of the librating streamlines evolves due to a combination of the planet radial motion and the history of magnetic torques acting on the librating material. Figure from \cite{2018MNRAS.477.4596M}.}
 \label{fig:vortensity_evolution}
\end{figure*}

 The important flow pattern in the coorbital region is similar to that of Type III migration, see Figure \ref{fig:flow}. In contrast to Type III migration, however, we are now dealing with low-mass planets that do not open up a partial gap that drives Type III migration \citep{2003ApJ...588..494M}. However, horseshoe drag for low-mass planets is, in an isothermal disk, driven by a \emph{vortensity} contrast \citep{2009ApJ...703..845C}, where vortensity is defined as the vertical component of the vorticity divided by the surface density:
 \begin{align}
\xi = \frac{\left.\nabla\times {\bf v}\right|_z}{\Sigma}.
\end{align}
All we need for a nonzero corotation torque is a contrast between the vortensity in the coorbital region and the unperturbed local disk. Then the two streamlines indicated in Figure \ref{fig:flow} will carry a different value of the vortensity, yielding a net torque on the planet \citep{2014MNRAS.444.2031P}. The resulting vortensity distribution is illustrated in Figure \ref{fig:vortensity_evolution}, where the left panel shows a disk where viscosity is strong enough to maintain the original vortensity gradient across the corotation region, and the right panel shows the inviscid case, where a vortensity contrast has been built up through a combination of migration and radial disk flow \citep{2018MNRAS.477.4596M}. 

The criterion for a vortensity contrast between the coorbital material and the rest of the disk to be maintained is that any diffusion should take place on longer timescales than the libration timescale \citep{2014MNRAS.444.2031P}. Therefore, the dividing line between the realm of dynamical corotation torques and viscous Type I migration is exactly the criterion for corotation torques to become saturated. If saturation happens, this means that dynamical torques enter the picture. 

One important difference between Type III migration and migration driven by dynamical torques is the way the vortensity or density contrast can be achieved. For high-mass planets, one can rely on the planet clearing a partial gap \citep{2003ApJ...588..494M}, or the planet can be placed on a very steep density gradient \citep{2008MNRAS.386..164P,2008MNRAS.386..179P,2008MNRAS.387.1063P}. While the latter option is available for low-mass planets as well, migration itself is capable of generating a vortensity contrast. Since the planet migrates together with all coorbital material, in the absence of diffusion the vortensity associated with the coorbital material will be the same as the disk vortensity \emph{at the starting location of the planet}. Therefore, if the planet is migrating in a disk with a radial gradient in vortensity, a contrast will build up automatically. The resulting dynamical corotation torque can either enhance or slow down migration.

If the migration direction of the planet (driven by the \emph{total} torque) is the same as the direction in which the corotation torque alone would push the planet, the planet will be pushed into a runaway migration, much as in the case of Type III migration. If, on the other hand, the migration of the planet goes against the will of the corotation torque, migration will slowly grind to a halt \citep{2014MNRAS.444.2031P}. Adding some viscosity leads to residual migration, driven by how much angular momentum is able to diffuse into the coorbital region. 
 
The picture can be generalised to non-isothermal or radiative disks. In this case, the thermal memory of the libration island gives rise to new migration regimes at high optical depths \citep{2015MNRAS.454.2003P, 2016MNRAS.462.4130P}.

Another generalization comes in the form of a magnetically braked disk \citep{2017MNRAS.472.1565M}. The poorly ionized planet-forming regions may well be virtually free of magnetic turbulence, except for a narrow layer near the top from which a wind is launched \citep{2013ApJ...769...76B,2015ApJ...801...84G}. If also the Hall effect is active, this creates significant laminar horizontal magnetic fields and associated Maxwell stress \citep{2013ApJ...769...76B, 2014A&A...566A..56L,2017A&A...600A..75B}. This stress generates a radial flow in the disk, and while a radial flow is expected in viscous accretion disks as well, in the laminar magnetic case there is no associated turbulence or diffusion. This means that even non-migrating planets would experience dynamical corotation torques, as the radial disk flow sets up the vortensity difference between the coorbital region and the rest of the disk. Depending on the ratio of the natural migration rate of the planet and the speed of the laminar disk flow, dynamical torques can either force the planet to migrate at the same speed as the disk flow, or cause a runaway \citep{2017MNRAS.472.1565M, 2018MNRAS.477.4596M}. It is worth noting that an extension to higher masses has been shown to work \citep{2020A&A...633A...4K}.

Above studies were mostly done in two spatial dimensions, working with vertically averaged quantities. While many of the features of horseshoe drag carry over in three dimensions \citep{2016ApJ...817...19M,2017AJ....153..124F}, there are important differences. One is that while migration appears to be insensitive to surface winds, in radiative disks an additional negative torque arises that is associated with buoyancy waves \citep{2012ApJ...758L..42Z,2020MNRAS.493.4382M}.  

\subsection{Intermediate Mass Planets and No `Inertial Limit' to Planet Migration}
\label{sec:inertial}

Increasing the planet mass while keeping the turbulent viscosity small brings us through the 'Feedback Mass' boundary in Figure \ref{fig:migration_org_chart}. This is where planets are massive enough to significantly alter the density profile in their vicinity.

The emerging new paradigm where protoplanetary disks are largely laminar with almost no turbulent diffusion \citep{2013ApJ...769...76B,2015ApJ...801...84G} has sparked renewed interest in the concept of the 'inertial limit' of planet migration \citep{1989ApJ...347..490W,1984Icar...60...29H,2002ApJ...572..566R}. In 1D models of laminar disks, \emph{all} planets will open up deep gaps unless they can migrate through the disk fast enough \citep[see also][]{2015ApJ...802...56M,2017Icar..285..145C}. For increasing planet mass, the density perturbation induced by the planet provides a negative feedback and migration stalls at the 'inertial limit' \citep{1989ApJ...347..490W}. The root cause of this stalling is the asymmetry in the induced density perturbation in the inner and outer disks.

Early explorations using multidimensional hydrodynamic simulations did indeed find that migration would stall or slow down near the inertial limit \citep{2009ApJ...690L..52L,2010ApJ...712..198Y}. \cite{2017ApJ...839..100F} found that super-Earths in inviscid disks could stall their migration, while still driving accretion\index{Accretion} onto the central star through the action of their density waves \citep{2001ApJ...552..793G}. Migration stalling in combination with pebble accretion can potentially explain the dichotomy between inner super-Earths and outer Gas Giants \citep{2018ApJ...859..126F}. 

\begin{figure}[h]
 \epsscale{0.8}
 \plotone{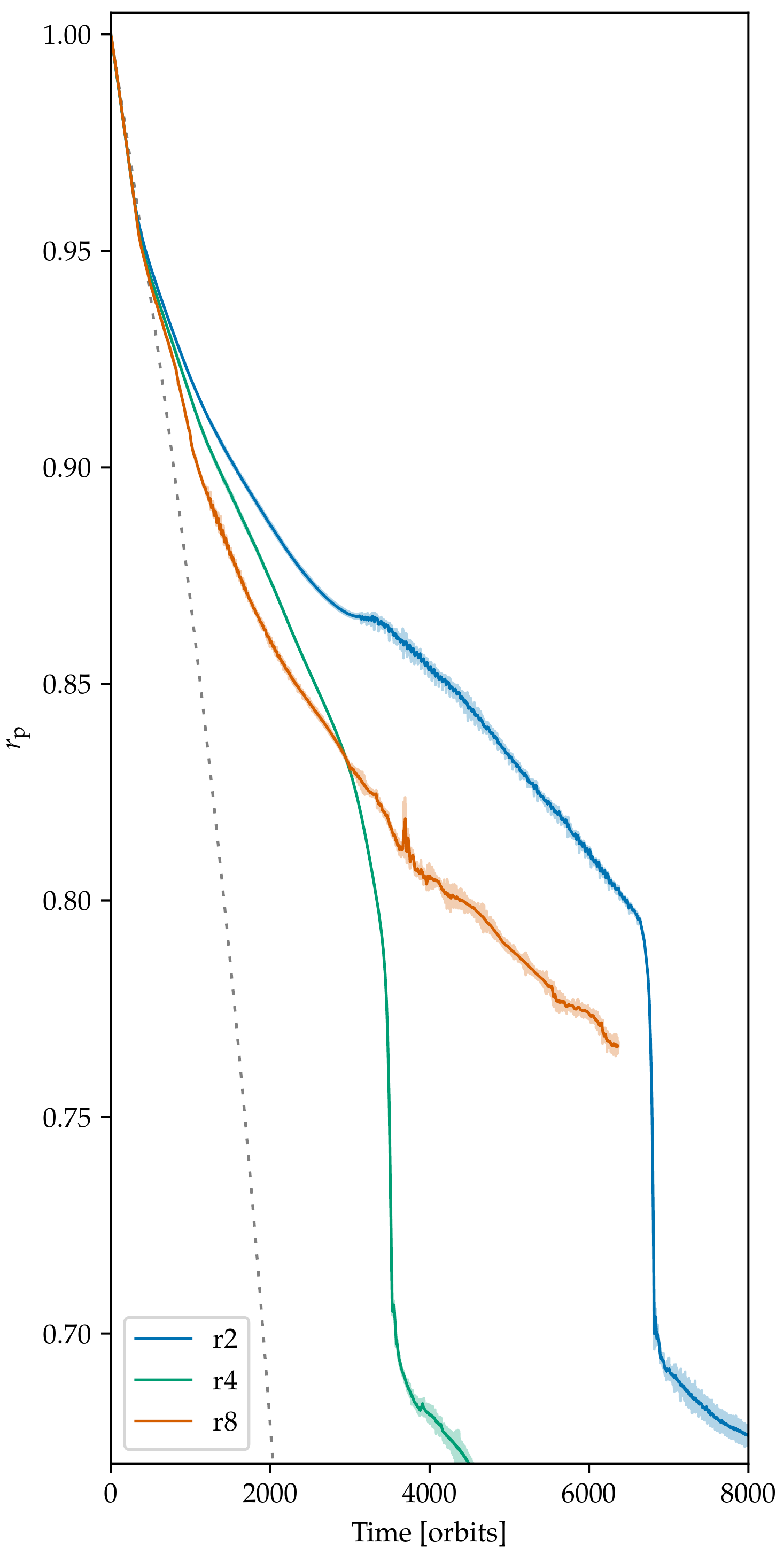}
 \caption{\small Migration of a $q=1.25\times 10^{-5}$ planet in an inviscid disk with $H/r=0.035$. The three resolutions have $23$ (r2), $47$ (r4) and $93$ (r8) grid cells per scale height at $r=r_\mathrm{p}$. Figure from \citet{2019MNRAS.484..728M}}
 \label{fig:feedback}
\end{figure}

disks with low levels of viscosity come with their own set of challenges. It has been known that intermediate-mass planets can excite vortices at the edges of their partial gap \citep{2003ApJ...596L..91K}. It was found in \cite{2019MNRAS.484..728M} that such vortices can seriously disrupt the picture of the inertial limit. An example is shown in Figure \ref{fig:feedback}. At the lowest resolution, r2, which has 23 zones per scale height, migration halts at around $3000$ orbits, which has been seen before in e.g. \cite{2010ApJ...712..198Y}. However, soon after this, vortices emerge around the orbit of the planet, and the interaction between the planet and the vortices pushes the planet back on an inward trajectory. Eventually, at around $7000$ orbits, it enters a phase of rapid, inward Type III migration. While the average migration rate is slow compared to viscous Type I (indicated by the dotted curve), migration is inward without any prolonged episodes of stopping. Moreover, doubling the resolution (r4) gives different results, with an early Type III phase that makes the average migration rate only about a factor of 2 slower than viscous Type I. Doubling the resolution again leads again to a different trajectory, now without any Type III episodes. This indicates that numerical convergence in these types of inviscid simulations is hard to achieve \citep{2019MNRAS.484..728M}.

Towards larger values of the viscosity, we cross the 'viscous smoothing' border in Figure \ref{fig:migration_org_chart} towards the realm of smooth feedback. This is where viscosity can partly smooth out the action of vortices, while feedback effects are still important. This typically happens at $\nu \sim 10^{-7}~r_\mathrm{p}^2\Omega_\mathrm{p}$ (corresponding to $\alpha \sim 10^{-5}$ for typical values of $H/r$) \citep{2010ApJ...712..198Y, 2019MNRAS.484..728M}. Migration in this regime can be very slow compared to viscous Type I migration.

\subsection{The Nature of Type II Migration in Viscous Disks\index{Disk!gaps}\index{Planet migration!Type II}}
%\label{sec:typeII}

\label{sec:nature_of_typeII}
The classical Type II migration rate assumes an empty planetary gap with no crossing flow. 
With an idealized deep gap, the planet is locked between the outer and the inner disks
and must migrate together with the radial viscous evolution of the disk. 
However, since around the turn of the century, many hydro-dynamical simulations had found significant gap-crossing flows for deep gaps formed by Jupiter-mass planets, as mentioned in Section \ref{sec:typeII}.
After PPVI, highly accurate hydrodynamical simulations have revealed structure of the planetary gaps and Type II migration rates for very wide parameter ranges of the planet mass $M_{\mathrm{p}}$, 
the disk viscosity $\nu$, and the disk scale height $H$. Owing to the reliable numerical results, a new qualitative and quantitative picture of the gap structure and Type II migration have been constructed.  This recent progress is reviewed below.

\subsubsection{Gap Depth and Gap Profile}
Structures of almost axisymmetric gaps are described by radial profiles of the azimuthally averaged surface densities of disks. One of the most important quantities characterizing a radial profile is 
the minimum surface density in the gap, $\Sigma_{\mathrm{gap}}$, or the ratio of $\Sigma_{\mathrm{gap}}$ 
to the unperturbed value, $\Sigma_0$, which is called the gap depth. Usually, the radial profile attains its minimum surface density at the planetary orbital radius, $r_{\mathrm{p}}$ if the high-density region around the planet is excluded at the density evaluation. Another important quantity for comparison to observations is the gap width.

An intensive survey of accurate hydrodynamical 
simulations on the gap structure was first carried out by \citet{2013ApJ...769...41D} for wide parameter ranges of $M_{\mathrm{p}}$ and $\nu$.   Their simulations with analytical considerations indicate that the gap depth is given by
\begin{align}
\frac{\Sigma_{\mathrm{gap}}}{\Sigma_0} = (1 + K/29 )^{-1}, 
\label{eq:gap_depth}
\end{align}
where $K = q^2/(h^5 \alpha)$
is a non-dimensional parameter \citep[][]{2015MNRAS.448..994K}. This empirical formula 
for the gap depth is consistent with the numerical results by \citet{2004ApJ...612.1152V}.
The formula was also confirmed by later surveys of hydrodynamical simulations and explained by zero- and one-dimensional analytic models \citep[e.g.][]{2014ApJ...782...88F,2015MNRAS.448..994K,kanagawa17deepgaps}.
The gap opening condition (\ref{eq:gap_opening_condition}) is consistent with equation~(\ref{eq:gap_depth})
because it is rewritten as $\Sigma_{\mathrm{gap}}/\Sigma_0 \lesssim 0.5$, using equation~(\ref{eq:gap_depth}).

\begin{figure}[h]
% \epsscale{0.8}
 \plotone{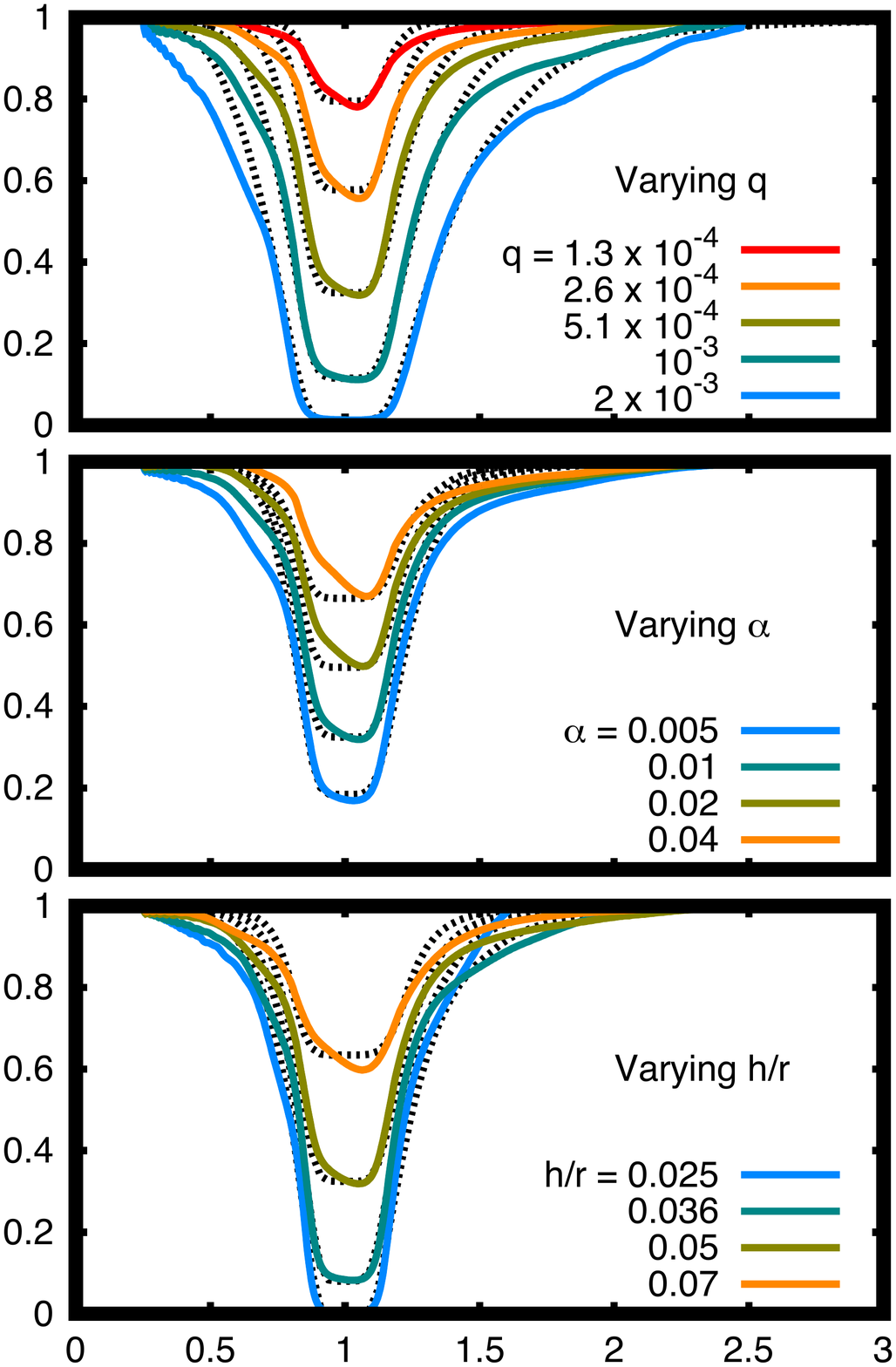}
 \caption{\small Comparison between the gap model of \cite{2020ApJ...889...16D} and a wide range of numerically calculated gaps.  Black dashed curves are the analytical model, and colored solid curves are the numerically calculated gap profiles in steady-state disks.  The fiducial disk model used $q = 5.1 \times 10^{-4}$, $\alpha = 0.01$, $h = 0.05$.  Each panel picks one of these parameters and varies it, keeping the others fixed at their fiducial values. Figure from \citet{2020ApJ...889...16D}}
 \label{fig:gaps}
\end{figure}

Some studies proposed more accurate models of the gap depth than equation~(\ref{eq:gap_depth}) for deep gaps of $K > 10^3$ 
\citep[e.g.][]{2014ApJ...782...88F,2018MNRAS.479.1986G,2020ApJ...889...16D}.  It is useful to introduce another gap surface density, $\Sigma_{\mathrm{gap,av}}$, which is 
radially averaged over the radial range between $r_\pm \equiv r \pm 2R_{\mathrm{Hill}}$.  The radially averaged surface density is physically more meaningful 
than $\Sigma_{\mathrm{gap}}$ ($\simeq \Sigma(r_{\mathrm{p}})$), 
because the planet interacts mainly with the gas at $r_\pm$ rather than at $r_{\mathrm{p}}$.
Equation~(\ref{eq:gap_depth}) has been found to be valid even for $K>10^3$,
if it is used for $\Sigma_{\mathrm{gap,av}}$ instead of $\Sigma_{\mathrm{gap}}$.
Accurate empirical models for the gap width and the radial profile of $\Sigma(r)$ have 
also been constructed
\citep[e.g.][]{duffell15gap,2020ApJ...889...16D,kanagawa16width,kanagawa17deepgaps}, including public tools available for download to compute the complete density profile $\Sigma(r)$ for given planet and disk parameters \citep[][see Figure \ref{fig:gaps}]{2020ApJ...889...16D}.

For very massive planets with $q \gtrsim 3 \times 10^{-3}$ or for disks with a low viscosity, 
structures of the deep gaps are unstable and the disks become eccentric
\citep[e.g.][]{2003ApJ...596L..91K,2006MNRAS.368.1123G,2006A&A...447..369K,2014ApJ...782...88F}.
The above gap models are not applicable to such eccentric disks.

\subsubsection{Type II Migration Rate in Viscous Disks}
The classical Type II migration rate is given by 
\citep[e.g.][]{1986ApJ...309..846L,1995MNRAS.277..758S}
\begin{align}
v_{r,\mathrm{classical}} =
\left\{
\begin{array}{ll}
\displaystyle -\frac{3\nu\,}{\,2r_{\mathrm{p}}} & \:\: (M_{\mathrm{p}} < A \Sigma_0 r_{\mathrm{p}}^2), \\
\displaystyle -\frac{3\nu\,}{\,2 r_{\mathrm{p}}} \frac{A \Sigma_0 r_{\mathrm{p}}^2}{M_{\mathrm{p}}}
& \:\: (M_{\mathrm{p}} \geq A \Sigma_0 r_{\mathrm{p}}^2).
\end{array}
\right.
\label{eq:classical_typeII_rate}
\end{align}
In this equation, the former is called the disk-dominated case and the later is 
the planet-dominated case. The constant $A$ is chosen to be 2-10, depending on 
the authors \citep[e.g.][]{2005A&A...434..343A,2014prpl.conf..667B}.

To verify the classical migration model, many hydro-dynamical simulations measured 
Type II migration rates, varying the disk parameters and the planet mass
\citep[e.g.][]{2014ApJ...792L..10D, 2015A&A...574A..52D, 2018ApJ...861..140K, 2018A&A...617A..98R}.
Those results suggest that the migration rate can vary as $v_r \propto \nu \Sigma_0 /M_{\mathrm{p}}$
for relatively massive planets \citep[although an anomalous additional dependence on $h$ may still be present, e.g.][]{2014ApJ...792L..10D}. 
This dependence is reasonably consistent with the planet-dominated case of the classical model. 

However, several points were found to be starkly inconsistent with the classical model.
For a relatively massive disk, a gap-forming planet can migrate faster than the 
radial flow velocity of the viscous accretion disk, $3\nu/(2r_{\mathrm{p}})$. 
The observed migration rate also depends on the disk aspect ratio $h$ for 
a fixed $\nu$. These results cannot be explained by the classical model.
\citet{2014ApJ...792L..10D} showed that no pile-up of the outer disk or no depletion 
of the inner disk occurs.  This means that a considerable gap-crossing flow exists.

A new picture of Type II migration proposed by \citet{2018ApJ...861..140K} appears to be much more consistent with numerical results. In this picture, the gap-forming planet migrates 
with the Type I torque determined by the surface density at the gap bottom instead of the unperturbed one.  Then, using equations (\ref{eq:torque_on_planet}) and (\ref{eq:gap_depth}), the migration rate is given by
\begin{align}
v_r 
&= \displaystyle 
2 C r_{\mathrm{p}}\Omega_{\mathrm{p}}
\frac{q^2}{h^2}  \frac{\Sigma_{\mathrm{gap}}r_{\mathrm{p}}^2}{ M_{\mathrm{p}}} \nonumber \\[2mm]
& \displaystyle \simeq - 150 h \frac{\nu\,}{\,r_{\mathrm{p}}}
\frac{\Sigma_0 r_{\mathrm{p}}^2}{M_{\mathrm{p}}}, 
\label{eq:new_typeII_rate}
\end{align}
where the gap opening condition (\ref{eq:gap_opening_condition}) (i.e. a very large $K$) is assumed and $C$ is set to be $\sim -3$ 
at the second equality. The new migration rate (\ref{eq:new_typeII_rate})
resembles that of the classical migration rate in the planet-dominated case, except the dependency on the aspect ratio $h$. 
This new model of the migration rate can also describe the continuous transition from the Type I migration to Type II.

The new migration model has been used in population synthesis calculations for giant planets 
\citep[e.g.][]{2018ApJ...864...77I,2019A&A...622A.202J,2021ApJ...914..102C}.
Note that the new Type II migration rate is 
further reduced by a rapid gas accretion\index{Accretion} onto the planet 
\citep[][]{2016ApJ...823...48T,2018A&A...617A..98R,2020ApJ...891..143T, 2020ApJ...891..166W}.
When rapid gas accretion onto the planet is regulated by a relatively low disk accretion rate, 
the planetary accretion further reduces the surface density in the gap, $\Sigma_{\mathrm{gap}}$.
Thus the short supply reduces equitably the planetary migration rate and accretion rate since both are proportional to $\Sigma_{\mathrm{gap}}$
\citep[][]{2020ApJ...891..143T}. This additional slow-down of Type II migration should also be included in population synthesis calculations.

Further attempts have been made to refine the new Type II migration model. \citet{2019arXiv190802326D} demonstrated 
that the back reaction of the net torque causes imbalance in the surface density between the outer and inner disk, 
which alters the torque imbalance (i.e. the net torque) on the planet in turn. They showed that this positive feedback 
enhances the net torque particularly in the low-viscosity case.
The torque imbalance due to the outer gap edge heating by the central star was examined by
two- and three-dimensional simulations \citep[][]{2018MNRAS.481.1667H, 2020A&A...642A.219C}.  Further debate on the new Type II migration model is ongoing 
\citep[][]{2020MNRAS.492.1318S, 2020arXiv200404335K}.

One element still missing from existing Type II migration models is the allowance for {\em outward} migration, which can occur due to a number of mechanisms.  Migration (and growth) of super massive planets with $q \gtrsim 10^{-2}$ (transitioning from a planet to a binary) was investigated by \citet{2020ApJ...901...25D}. They found outward migration for $q>0.05$ as well as significant accretion onto the secondary, always driving the system toward an equal-mass binary. Similar outward migration of supermassive planets was reported by \citet{2021arXiv210505277D}. They posited that the outward migration is due to the unstable eccentric outer disk. Interestingly, outward migration is not predicted by classical Type II migration, nor is it predicted by the new migration formulas of \cite{2018ApJ...861..140K}. This by itself suggests there is substantial room for improvement of analytical models. In \cite{2020ApJ...900...44C}, the authors investigated the effect of the changing location of Lindblad resonances due to the gap profile and concluded that massive planets that open up deep gaps may indeed migrate outward.

For intermediate planet masses, other effects can conspire to yield, at least temporarily, outward migration. In disks with very low viscosity, super-Earths of $q \sim 10^{-5}$ can open up gaps that are unstable to the Rossby Wave Instability \citep{1999ApJ...513..805L}. The interaction of the planet with the resulting vortices can drive outward migration \citep{2019MNRAS.484..728M}, similar to what was seen for Saturn-mass planets \citep{2010MNRAS.405.1473L}. Outward migration of super-Earths was also seen by \cite{2015ApJ...806..182D} in disks of moderate viscosity.

\subsection{Impact of Embedded Solids on Planet Migration\index{Dust!Gas drag}}
Most work so far has concerned the gas component of the disk only. While the dust component is recognized to be important for observations \citep[e.g.][see also section \ref{sec:obs}]{2004A&A...425L...9P}, its effect on disk planet interactions has largely been ignored due to the small dust mass compared to the gas. Recent work, however, has recognized that the solid component can produce two new torque effects on the planets, the dust scattering torque \citep{2018ApJ...855L..28B, 2018MNRAS.478.2737C} and aero-resonant migration \citep{2019MNRAS.490.1861S}. 

The dust scattering torque arises from dust particles that stream past the planet. The resulting torque bears resemblance to Type III migration, where the dust particles follow streamlines as depicted in Figure \ref{fig:typeiii}. The strength of the Type III torque is such that it can often compensate for the small dust mass compared to the gas \citep{2018ApJ...855L..28B}. This effect is most prominent for larger particles that are not very well coupled to the gas. \cite{2018MNRAS.478.2737C} considered smaller particles, and found oscillatory torques from the coorbital region for high dust-to-gas ratios, together with vortex formation. In the thermodynamic framework of \cite{2017ApJ...849..129L}, the vortices can be attributed to baroclinic effects stemming from a variation in dust-to-gas ratio \citep{2018MNRAS.478.2737C}. More recently, \cite{2020MNRAS.497.2425H} found that a dusty version of dynamical corotation torques (see section \ref{sec:dynamical}) can significantly slow down migration.

Aero-resonant migration \citep{2019MNRAS.490.1861S} involves solid bodies of sizes between 10 m and 10 km. When these drift inward towards a planet due to gas drag, they get caught in mean motion resonances. Here, they can exchange angular momentum with the planet and drive inward migration. The resulting migration rate can be comparable or larger than the viscous Type I migration rate, in particular in the inner disk \citep{2019MNRAS.490.1861S}, if the total mass of planetesimals captured in mean motion resonance consists of a significant fraction of the mass of the planet.  

Gap-opening planets were considered by \cite{2019ApJ...879L..19K}, who found that dust feedback on the gas decreased the gas surface density at the outer edge of the gap, thereby reducing the torque on the planet. For wide and deep gaps, the direction of migration can reverse due to dust feedback \citep{2019ApJ...879L..19K}.

\subsection{Migration in Wind-driven Protoplanetary Disks}
\label{sec:wind}

The realization that protoplanetary disks could be largely laminar, with accretion\index{Accretion} driven by a wind rather than turbulent diffusion \citep{2013ApJ...769...76B,2015ApJ...801...84G} has led to new studies of disk-planet interactions in such disks. Consequences of the low level of turbulence have been discussed already in sections \ref{sec:dynamical} and \ref{sec:inertial}. Here, we focus on recent progress specific for wind-driven disks.

At low planet masses, the surface accretion flow driven by a wind is likely largely decoupled from the planet migration \citep{2020MNRAS.493.4382M}. Any surface accretion flows are too high up in the disk, where gas densities are too low, to influence the migration speed of the planet. No vertical communication between the accreting layers and the midplane was observed in 3D simulations \citep{2020MNRAS.493.4382M}.

Most studies of disk-planet interactions in wind-driven disks use a vertically integrated approach, where the effect of the disk wind is incorporated as an additional torque on the disk. \cite{2015A&A...579A..65O,2015A&A...584L...1O} incorporated the model of \cite{2010ApJ...718.1289S} into a one-dimensional viscous evolution model for the disk, and constructed migration maps based on the formulae of \cite{2011MNRAS.410..293P}. Type I migration was found to be strongly reduced for a wide range of wind parameters due to the modified surface density slope induced by the disk wind. Effects of rapid, wind-driven gas flow on the saturation of the corotation torque was investigated in \cite{2017A&A...608A..74O}. Systems of low-mass planets in wind-driven disks were found to undergo dynamical instabilities, thereby avoiding resonant configurations \citep{2018A&A...615A..63O}.

At higher planet masses, 2D simulations of higher-mass planets in wind-driven disks showed that dynamical torques play a role in this mass regime (Saturn-Jupiter) as well \citep{2020A&A...633A...4K}. Fast enough accretion flows can drive planets into Type III migration that can be directed outward \citep{2020A&A...633A...4K}.

\subsection{Migration of Planets Formed by Gravitational Instability\index{Disk fragmentation}\index{Disk!Fragmentation}\index{Disk!Gravitational instabilities}}
Gravitational Instabilities (GIs) may form planets from collapsing clumps in very early massive disks \citep{1951PNAS...37....1K,1978M&P....18....5C,1997Sci...276.1836B}. The physical regime is quite different from planet migration in a fiducial GI stable disk. It was studied in detail in \cite{2011MNRAS.416.1971B}, who found that newly formed fragments would migrate in the Type I regime, despite being Jupiter-sized. This is because the time scale of GI to form a clump is the orbital time scale, which is much shorter than the gap formation time scale. The planet therefore has no time to open up a gap and starts Type I migration until it slows down enough to open a gap \citep{2011MNRAS.416.1971B,
2015ApJ...802...56M}. Interaction with global spiral modes can slow down or even halt migration \citep{2011ApJ...737L..42M}. 

Studies with more realistic thermodynamics in the form of radiative disks were presented in \cite{2012ApJ...746..110Z}, \cite{2015ApJ...810L..11S}, \cite{2018A&A...618A...7V} and \cite{2018MNRAS.477.3110S}. They found that after an initial phase of fast migration, planets are able to open up gaps and slow down or even reverse the direction of migration. The radiative feedback of accretion onto the planet is important in determining whether the planet can survive at large orbital distances \citep{2015ApJ...810L..11S}. Recently, the importance of thermodynamics was further emphasized in \cite{2020MNRAS.496.1598R}, who found that migration halts if the inner parts of the disk are gravitationally stable. Subsequent tidal and thermal evolution of clumps can change their mass, size and dust content, such that the resulting planet population could show a great deal of diversity \citep[see e.g.][]{2017PASA...34....2N}.

A code benchmarking study on planet migration and growth in self-gravitating disks was presented in \citet{2019MNRAS.486.4398F}. Migration rates between the various codes varied between $10\%$ and $50\%$, depending on the numerical setup. In particular, they pointed out the importance of the artificial viscosity prescription and the way accretion was handled in the simulation in determining the migration outcome.

\subsection{Migration near the inner disk edge\index{Disk!Inner hole}}

Here we discuss recent progress on disk-planet interactions near the inner edge of the protoplanetary disk. If the disk is truncated by the stellar magnetic field, there will be a low density inner magnetospheric cavity connected to the bulk of the disk. The steep, positive density gradient at the edge of the cavity can have important effects on migration. 

If the width of the cavity edge is smaller than the width of the horseshoe region, an approximate expression for the corotation torque is the \emph{one sided} horseshoe drag \citep{2017A&A...601A..15L}. The resulting outward migration as the cavity expands is able to break up resonances formed by earlier convergent migration \citep['magnetospheric rebound',][]{2017A&A...601A..15L,2017A&A...606A..66L}.

The question if and where planets are stopped near the inner edge has received a lot of attention. Wave reflection off sharp edges was known to affect the Type I torque \citep{2011ApJ...741..109T}, and it was shown in \cite{2018MNRAS.473.5267M} that this can lead to the planet stopping as far out as 3 times the radius of the inner edge. It was argued in \cite{2019MNRAS.485.2666R} that disks may have several edges where planets could stop migrating. The migration of a resonant chain of planets into the cavity was studied in \cite{2021A&A...648A..69A}, who found that even chains of planets are efficiently stopped, and found that overstable librations \citep[see][]{2014AJ....147...32G} may give rise to compact systems. \cite{2019A&A...630A.147F} studied the effect of the silicate sublimation front on the location where planets halt their migration, and found that Earth and super-Earth planets stop at periods ranging from 10 to 22 days, in good agreement with observations.

Near the inner disk edge is also the place where star-planet tidal interactions might play a major role in driving orbital evolution. In particular, planets that stall their migration near the inner disk edge may undergo further tidal migration once the disk has disappeared. For an overview of star-planet tidal interactions we refer the reader to \cite{2014ARA&A..52..171O,2020oeps.book..191O}.
%\cite{2020arXiv200903090R}

\subsection{Eccentricity and Inclination Evolution\index{Disks!Eccentricity damping}}
\label{sec:eccentr-incl-evol}

Planet--disk interactions alter the planet's orbital eccentricity and inclination as well as its semimajor axis, and their evolution should be treated consistently \citep{ida20}. For planets that are unable to open a gap in the disk, small eccentricities and inclinations $e,i\la h$ lead to small horizontal and vertical excursions within the disk. Until recently, it was believed that such $e$ and $i$ would be damped on a timescale much shorter than that of Type-I migration, owing to the emission of density and bending waves in the surrounding disk \citep{tanaka04}. However, different conclusions have been reached by recent numerical simulations \citep{2017MNRAS.469..206E,2017A&A...606A.114C} and linear analysis \citep{2019MNRAS.485.5035F} that include radiative thermal diffusion and treat the accreting planetary embryo as a heat source that produces a hot, underdense trail in the disk (see Section~\ref{sec:thermal}), as the associated forces are typically larger than those due to the emission of waves for planets of mass comparable to or lower than the critical mass of Eq.~(\ref{eq:masscrit}). If the planet's luminosity is below the critical value of Eq.~(\ref{eq:lumcrit}), then $e$ and $i$ are damped (and typically much faster than in an isothermal or adiabatic disk); above the critical luminosity, both $e$ and $i$ grow to values comparable to $h$. The growth of $e$ is significantly faster and tends to dominate the evolution, and the effect appears to be suppressed when the planet exceeds a few Earth masses.

It has been appreciated for some time that there is a qualitative change in behaviour when $e\gtrsim h$, as this leads to supersonic relative motion between the planet and the disk. \citet{ida20} have proposed a simple prescription, based on dynamical friction calculations, for the damping of $a$, $e$ and $i$ of low-mass planets, that smoothly spans the sub- and supersonic regimes. %\citet{2021ApJ...915..113B} have made a detailed numerical study of the 3D, time-dependent flows around and onto planets of up to twice the thermal mass with eccentricities of up to $0.1$, in the isothermal case.

For planets of about one Jupiter mass or above, capable of opening a gap in the disk, the evolution of $e$ and $i$ over wide ranges of those quantities has been measured in numerical simulations by \citet{2013A&A...555A.124B}, omitting the thermal effects discussed above. Generally $e$ and $i$ are both damped,  although the damping timescales are greatly increased when these quantities are larger and the planet moves supersonically with respect to the disk. The measured damping rates have been compared with theoretical expectations and used to explain observed systems \citep{2017A&A...598A..70S}.
Important exceptions to this behaviour are found (i) when $e$ and $i$ are both small and the planet is sufficiently massive to open a deep gap, in which case $e$ grows; (ii) when $i\ga 40^\circ$, in which case there can be non-monotonic behaviour related to Lidov--Kozai cycles.

Earlier work on the eccentricity evolution of planets that open deep gaps \citep{goldreich03} had suggested that $e$ grows as a result of eccentric Lindblad resonances\index{Disks!resonances}\index{Lindblad resonances}\index{Planet formation!Resonances} when $e$ exceeds a small critical value, such that the opposing eccentric corotation resonances are saturated \citep{ogilvie03}. \citet{2015ApJ...812...94D} measured $\dot e>0$ in 2D numerical simulations for an intermediate range of $e$ ($0.04\la e\le 0.07$ for $q=10^{-3}$), attributing the upper limit to the collision of the planet with the disk. Eccentricity growth requires a clean gap to eliminate the coorbital Lindblad resonances, and is possible when the dimensionless parameter $K=q^2/\alpha h^5$ exceeds a threshold of $10^3$--$10^4$. \citet{2018MNRAS.474.4460R} simulated cases with $q=1.3\times10^{-2}$ and with two disk:planet mass ratios over $300000$ orbits, finding that both planet and disk become eccentric. Differently from \citet{2015ApJ...812...94D}, they found that no seed eccentricity is required, and that $e$ can grow to a value significantly larger than $h$. In 2D simulations with $q=10^{-3}$ and very low viscosity, \citet{lega21} found excitation of $e$ up to a much higher value of $0.25$, but the same did not occur in 3D simulations.

Since gap-opening planets often have comparable angular momenta to the disks in which they orbit, and since $e$ and $i$ are exchanged between the planet and the disk through secular gravitational interactions, it can be important to account for the coupled motions of the planets and disks, taking into account the mean-motion resonances that contribute to excitation and damping of $e$ and $i$. \citet{2016MNRAS.458.3221T} formulated a linear theory of this interaction for small $e$, highlighting differences in the behaviour of $e$ in 2D and 3D disk models. This type of analysis can explain the growth of disk eccentricity for planets above a few Jupiter masses \citep{2017MNRAS.467.4577T}, and \citet{2018MNRAS.474.4460R} explained the long-term evolution of the eccentricities of the planet and disk in their simulations in terms of coupled modes of the system.

\citet{petrovich19} proposed that the conserved angular-momentum deficit (AMD) of a system of two giant planets could be transferred from the outer planet to the inner one, conferring it with a large eccentricity, through the dispersal of the protoplanetary disk. However, the effect may be suppressed if the disk is also allowed to acquire (and to damp) eccentricity \citep{2019MNRAS.490.4353T}.

If a planet migrates into a central cavity in the disk, then both the disk and the planet can become significantly eccentric as a result of the eccentric Lindblad resonances that remain in the massive part of the disk \citep{2016MNRAS.458.3221T}. \citet{2021MNRAS.500.1621D} found that $e$ as large as $0.4$ could be achieved in this way.

Several recent studies have addressed the inclination evolution of planets on circular orbits. \citet{2018MNRAS.475.3201A} measured the effect of non-zero $i$ on the inclination damping rate and migration rate of low-mass ($q=10^{-4}$) planets, finding that both increase for small non-zero $i$, but are greatly reduced for large $i$. \citet{2018MNRAS.481...20N} and \citet{zhu19} showed further that planets that open a deep gap break the disk into independently precessing inner and outer disks, resulting in a warped shape.

In conclusion, it appears that there are several situations in which planet--disk interactions can cause a planet to acquire a non-zero orbital eccentricity. The inclusion of non-trivial thermal physics makes a significant difference to the outcome for low-mass planets. Since the eccentricity dynamics of gap-opening planets can be strongly coupled to that of the disk, it may also be important in that context to include the effects of heating and cooling, along with 3D dynamics.

\subsection{Multiple Planets}
This section reviews work on planet-disk interactions in the generalized case of multiple planets, restricted to work where the details of the planet-disk interaction is highlighted, including the formation of resonant chains. In particular the existence of resonant chains is an important observational diagnostic for planet-disk interactions \citep[see also][]{2014prpl.conf..667B}. Particularly well-known systems involving chains of mean motion resonances include Trappist-1 \citep{2017Natur.542..456G} and Kepler-223 \citep{2016Natur.533..509M}. In order to account for the diversity of orbital period ratios between adjacent pairs of planets, for example amongst the compact multi-planet systems discovered by Kepler, several studies in recent years have focused on how to break any resonant chains that arise during planet-disk interactions.

Dynamical instabilities are a prime candidate for taking planetary systems away from resonance. This idea was first presented in \cite{2012Icar..221..624M}, and subsequent work \citep[e.g.][]{2014A&A...569A..56C,2017MNRAS.470.1750I} showed that the distribution of planets as produced by  disk migration followed by dynamical instabilities can match the observations \citep{2021A&A...650A.152I}. 

\citet[][see also section \ref{sec:wind}]{2018A&A...615A..63O} studied the formation of close-in super-Earths in wind-driven disks. In cases where the surface density profile is such that Type I migration is significantly reduced, it was found that dynamical instabilities set in, destroying any resonant chains that Type I migration had set up. The resulting distribution of period and mass ratios can reproduce the observed distribution of close-in super-Earths \citep{2018A&A...615A..63O}.

For a pair of planets in both viscous and inviscid Type I migration, \cite{2018MNRAS.474.3998H} found that while convergent migration creates mean motion resonances between the planets, these can be destroyed by overstable librations, as suggested by \cite{2014AJ....147...32G}. This effect can potentially explain the relative lack of mean motion resonances in Kepler systems \citep{2018MNRAS.474.3998H}.

The role of viscosity in forming resonant chains was explored in 
\cite{2019MNRAS.489L..17M}. It was found that viscous disks indeed tend to form resonant chains, but that the picture changes qualitatively for inviscid disks. When the emergence of vortices (see section \ref{sec:inertial}) starts to make migration paths non-smooth, convergent migration into resonances is disrupted. Overall, migration is still directed inward, creating compact planetary systems that are not protected by resonances. This frequently leads to dynamical instabilities in addition to closely packed, stable non-resonant systems \citep{2019MNRAS.489L..17M}. 

Above studies indicate that while Type I migration almost inevitably leads to resonant chains of planets, this is by no means the end of the story. Both viscous and inviscid disks have plausible mechanisms for breaking the chains, either before or after the gas disk has disappeared. 

The role of the disk mass was explored both in \cite{2018A&A...611A..37M} and \cite{2020A&A...635A.204A}. \cite{2018A&A...611A..37M} studied the effect of a finite disk mass on the resonance location for a two-planet system (without considering disk migration) both for the 2:1 and the 3:2 mean motion resonance. \cite{2020A&A...635A.204A} focused on a particular numerical difficulty when dealing with migrating planets. When the disk self-gravity is ignored, there is an inconsistency lurking in the torque calculation: the planet feels and reacts to the gravity of the gas, but the gas does not feel its own gravity. Having planet and disk orbit in different gravitational potentials causes an error in the torque calculation \citep[see][]{2008ApJ...678..483B}. \cite{2020A&A...635A.204A} found that when this inconsistency is not corrected for, more compact planetary systems are produced.

\cite{2020ApJ...894...59K} focused on planets of higher mass, capable of opening a gap. Based on the empirical migration speed formula developed in \cite{2018ApJ...861..140K}, they derive a criterion for convergent migration of massive planets that compares favourably with hydrodynamic simulations. 

\subsection{Planets in binary star systems\index{Binary systems!planet formation}}

One area that was not discussed in the previous Protostars and Planets review \citep{2014prpl.conf..667B} concerns disk-planet interactions in binary star systems. Two basic configurations are possible: the planet can be in an S-type orbit, going round only one of the stars, or in a P-type orbit, where it orbits both stars. The latter are also known as \emph{circumbinary planets}.

Disk-planet interactions are relevant for two aspects of this three-body problem. First, while the disk is present, the planet will interact with either the circumprimary or the circumbinary disk\index{Circumbinary disks}. Second, any disk in the system will be tidally truncated by the companion star, in a similar way a planet opens up a gap in the disk. A disk in an S-type configuration will be truncated on the outside, while a circumbinary disk will have an inner hole. In addition, the binary companion can drive the gas disk to become eccentric \citep[e.g.][]{1991ApJ...381..259L,2008MNRAS.386..973P}, even if its own orbit is circular \citep{2008A&A...487..671K}. 

Early work focused mostly on the circumprimary case, and in particular the famous case of $\gamma$-Cephei \citep{2003ApJ...599.1383H}. Both Type 0 \citep{2008MNRAS.386..973P} and Type I/II migration \citep{2008A&A...486..617K} were investigated. Perhaps the most striking difference with the single star case is the excitation of eccentricity in the orbit of the planet, driven mostly by the eccentric gas disk \citep{Kley2010}.  

The discovery of Kepler 16-b by the Kepler Space Mission \citep{2011Sci...333.1602D} has spurred research into circumbinary planets. At the time of writing, roughly a dozen circumbinary planets are known \citep[for a recent overview, see][]{2018MNRAS.480.3800H,2021A&A...645A..68P}. In all cases, the orbital planes of the binary and the planet are well aligned. One striking feature of the circumbinary planet population is that they orbit close to the stability limit at roughly $3.5$ binary separations \citep{2021A&A...645A..68P}. Any closer and the orbit of the planet would be unstable \citep{1986A&A...167..379D}. At the same time, disk-planet interactions provide a natural stopping point for inward migration near the inner edge of the circumbinary cavity \citep{2008A&A...483..633P}. The final orbital parameters of the planet at the location where migration halts depends on the physical state of the disk (viscosity, mass, thermodynamics), making circumbinary planets ideal test locations for disk-planet interactions \citep{2013A&A...556A.134P,2014A&A...564A..72K,2015A&A...581A..20K,2017MNRAS.469.4504M, 2018A&A...616A..47T,2019A&A...627A..91K,2021A&A...645A..68P}. 

\subsection{Circumplanetary Disks\index{Giant planets!Formation}\index{Giant planets!Gas accretion}}
Forming planets are expected to gather material from the protoplanetary disks around themselves, forming their primordial atmospheres and potential breeding grounds of moons and satellites. Similar to the circumstellar disks around forming stars, this structure is conventionally named ``circumplanetary disk''(CPD), although, intriguingly, it may not be disk-like at all. In fact, whether the CPDs are truly bound to their host planets is also under contention, as we will see later in this section.

Simulating CPDs is numerically challenging because their small sizes demand resolutions much higher than typical simulations of planet-disk interaction---orders of magnitude higher. The length scale for CPD dynamics is either the Hill radius $r_{\rm H}$, or the Bondi radius $r_{\rm B}$, whichever is smaller. For Jupiter mass planets around solar-type stars, $r_{\rm H}$ is the smaller of the two and is about 7\% of its orbit's semi-major axis; for Earth-like planets at 1 au, $r_{\rm B}$ is staggeringly small, just about 0.3\% of their semi-major axes. To resolve a length scale so small requires special numerical treatment, such as grid refinement \citep{2003ApJ...586..540D,machida08,wang14,szulagyi16,2021ApJ...915..113B}, non-uniform grid geometry \citep{fung15cpd,Schulik20}, or local simulations \citep{ormel15}.

Simulations allow us to extract sizes and rotation rates of CPDs, as well as, in models with thermal cooling included, the rates of gas accretion onto the planets. We will first give an overview of these topics, and then review the latest findings. For CPD sizes, a common misconception is that they equal $r_{\rm H}/3$, which may have originated from an estimate made by \cite{Quillen98} through assuming the accreted gas has an incoming angular momentum of $r_{\rm H}^2 \Omega_{\rm p}$. Numerical works have reported a range of different sizes, showing that the CPD size has a more complex dependence on planet mass and that it is also sensitive to the thermal properties of the gas. Still, the value $r_{\rm H}/3$ may have some relevance, since it is roughly similar to the largest stable orbit for ballistic particles orbiting a planet, which \cite{martin2011} found to be $\sim0.4r_{\rm H}$. 

Like its size, the kinematics of a CPD is also dependent on both planet and gas properties. Moreover, it can only be accurately captured in 3D simulations. Pioneering work done using 2D simulations have found that gas flows onto the CPD near the L1 and L2 Lagrange points \citep{1999ApJ...526.1001L}, but later, 3D simulations reveal that it is a vertical flow near the poles of the planets that feed the CPDs \citep{2003ApJ...586..540D}. In fact, in the midplane, gas generally flows away from the planet, and the L1 and L2 Lagrange points are exit, rather than entry, points \citep{machida08,tanigawa12,ormel15,fung15cpd}. The incoming angular momentum of the gas, which determines the final spin of the CPD, is consequently much different between 2D and 3D. This leads to a large body of work aimed at understanding the 3D rotation of CPDs, and some converged results are starting to emerge. 

Measuring the gas accretion\index{Accretion} rate is a main goal of CPD studies, since it directly informs us about the formation of gas giants. Simulating accretion, however, remains highly challenging. To do so realistically requires modeling radiative cooling, but radiative transport is computationally expensive. Adding to that, planets are expected to cool slowly, taking millions of years to cool and become gas giants. 3D simulations that simultaneously solve the full transport equation, resolve the CPDs, and track their evolution for millions of years are simply far out of reach. Typical CPD simulations with radiative transport employ flux-limited diffusion and last for tens of orbits in global models \citep{ayliffe09,szulagyi16,szulagyi17,lambrechts17,lambrechts19}, or thousands of orbits in local models \citep{cimerman17}.

Under the local shearing sheet approximation, the equations governing the fluid dynamics around CPDs are greatly simplified. Ignoring viscosity and thermal dynamics, the entire set of Euler equations can be non-dimensionalized and be characterized by a single non-dimensional parameter \citep{1996ApJS..105..181K,machida08}. We call this parameter the ``thermal mass'' of the planet:
\begin{equation}
    q_{\rm th} = \frac{q}{h^3} \, .
\end{equation}
%JF: added text below
We distinguish it from the thermal mass of the {\it disk}, which equals the denominator on the right-hand side times the mass of the star (i.e., $h^3 M_{\star}$). The thermal mass of the planet is then the mass of the planet divided by the thermal mass of the disk.
It is straightforward to show that given a $q_{\rm th}$, the ratios between the key length scales $r_{\rm H}$, $r_{\rm B}$, and $H$ are all uniquely determined. In particular, $r_{\rm H}=r_{\rm B}$ when $q_{\rm th}=\sqrt{1/3} \sim 0.58$; $r_{\rm H}=H$ when $q_{\rm th}=3$; and $r_{\rm B}=H$ when $q_{\rm th}=1$. Hence, planets with $q_{\rm th}\gg1$ have $r_{\rm B}>r_{\rm H}>H$, and we call them ``super-thermal'' planets. The opposite limit is ``sub-thermal'' planets, which have $q_{\rm th}\ll1$ and $H>r_{\rm H}>r_{\rm B}$. This is a useful categorization that will help us identify trends in numerical results.

Another useful categorization is whether the models simulate isothermal gas or not. The isothermal assumption was commonly employed in early work, chosen for its simplicity, but mounting evidence suggests that non-isothermal gas behaves vastly differently in CPDs. In the following sections, we will separate our review into isothermal disks and non-isothermal disks.

\subsubsection{Isothermal CPDs}

\begin{figure*}[h]
 \includegraphics[width=\textwidth]{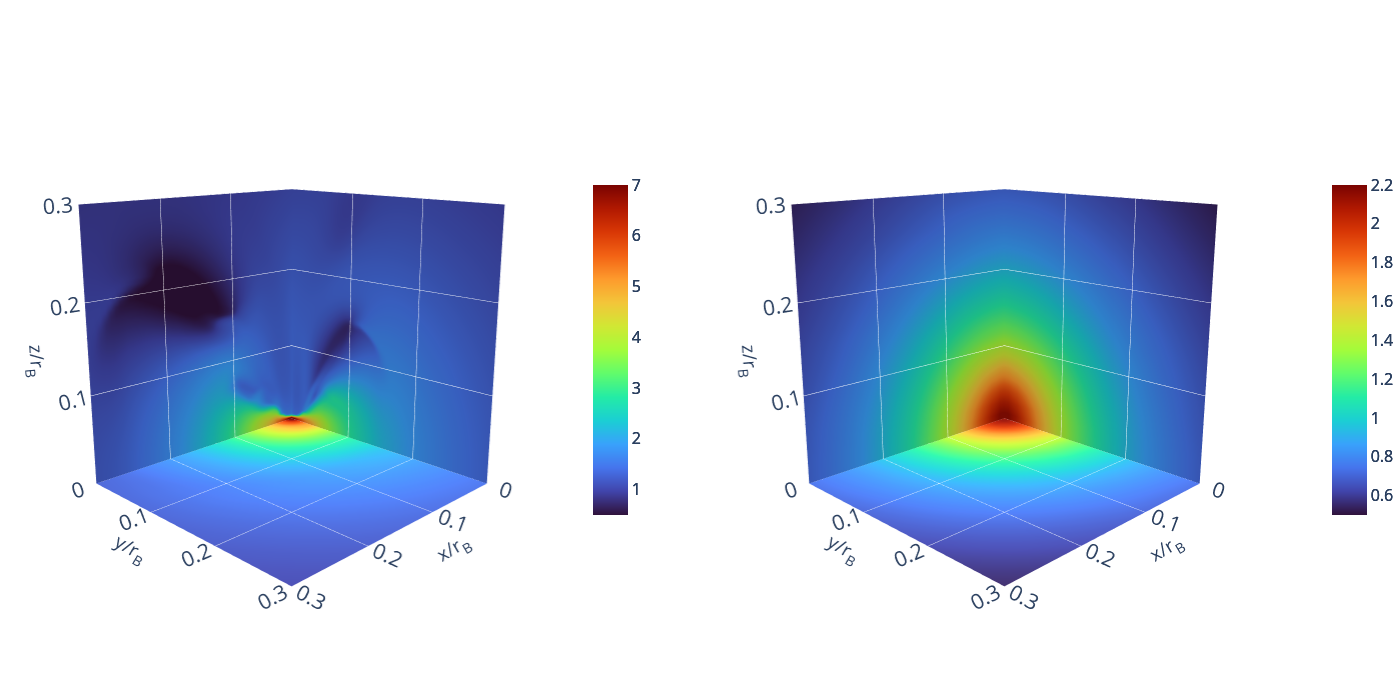}
 \caption{\small Simulations of CPDs around $q_{\rm th}=0.1$ planets taken from \cite{fung19cpd}. The plots show slices of density with the planet located at the origin. Density is in base-ten logarithmic scale, in units where the unperturbed density at the planet's location is 1. The isothermal simulation on the left shows a dense, flatten CPD, while the adiabatic one on the right is nearly spherically and orders of magnitude lower in density. Thermodynamics is critical in determining CPD structure.}
 \label{fig:CPD}
\end{figure*}

First, we review past findings on the rotation rate of isothermal CPDs. Early work by \cite{machida08,tanigawa12,wang14} found rotationally supported, Keplerian disks around their planets. However, later, simulations by \cite{ormel15}, \cite{fung15cpd} and \cite{kurokawa18} found CPDs that are slowly rotating or not at all. Soon after, \cite{fung19cpd,bethune19,2021ApJ...915..113B} found that isothermal CPDs are rotationally supported after all. The main reason for the apparent disagreement is that they simulated planets with different values of $q_{\rm th}$ and made different assumptions about the physical sizes of their planets. Upon closer inspection, the conclusions drawn by different groups are in fact consistent with one another.

Without accounting for the planet's physical size (i.e. a point mass), it is highly likely that CPDs are rotationally supported at short distances from their planets, for all values of $q_{\rm th}$. \cite{machida08} derived empirical relations showing that the specific angular momentum in isothermal CPDs, $l_{\rm CPD}$, scales with $q$ when $q_{\rm th}\lesssim 8$ and $q^{2/3}$ for larger $q_{\rm th}$. In agreement with their results, \cite{fung19cpd} found that $l_{\rm CPD} \sim 0.23 c_{\rm s} r_{\rm B}$ for sub-thermal planets. To better visualize this value, it is helpful to translate it to the ``centrifugal radius'' $r_{\rm c} = l_{\rm CPD}^2 / (GM_{\rm p})$, and in this case, we have $r_{\rm c}\sim 0.05 r_{\rm B}$. This measurement was consistent with the simulations by \cite{wang14}, and was later observed again by \cite{2021ApJ...915..113B}. On the super-thermal end, \cite{tanigawa12} measured $l_{\rm CPD} \sim 0.2 r_{\rm H}^{2} \Omega_{\rm p}$ for their $q_{\rm th}=3$ planet, which translates to $r_{\rm c}\sim 0.013 r_{\rm H}$. These measurements are in line with expectations in one sense---the disk size $r_{\rm c}$ does appear to scale with min$(r_{\rm H},\,r_{\rm B})$---but they are surprising in that the disks are very small, about one to two orders of magnitude smaller than min$(r_{\rm H},\,r_{\rm B})$. The left panel in Figure~\ref{fig:CPD} shows one example of these disks.

The fact that these disks are so small may be the reason why some isothermal simulations did not find rotationally supported CPDs. Numerical simulations of planet-disk interaction commonly include a smoothing length $r_{\rm s}$ in the gravitational potential of the planet to avoid the singularity of a point mass. One may interpret $r_{\rm s}$ as the physical size of the planet. \cite{ormel15}, \cite{fung15cpd} and \cite{kurokawa18} all reported little rotation in their simulations of sub-thermal planets, and, as it turns out, they all employed a $r_{\rm s}$ larger than $0.05 r_{\rm B}$. Similarly, \cite{bethune19} modeled the physical size of the planetary core as an inner boundary in local simulations, and found that the inner boundary has to be smaller than about $1/32\,r_{\rm B}$ in order to recover the Keplerian disk. In reality, the physical size of the planet relative to its Bondi or Hill radius depends on its orbit size; it can well be a significant fraction of $r_{\rm B}$ if the planet is close to its host star.

Next, we review measurements of the sizes of isothermal CPDs. This is a separate measurement from $r_{\rm c}$, which is the size of the rotationally supported disk. Beyond $r_{\rm c}$, the gas may be partially or completely supported by gas pressure, but remains bound to the planet and therefore should be considered a part of the CPD. On this topic, there is currently no clear consensus on the definition of where the CPD ends and the circumstellar disk begins. An intuitive approach may be to inspect the velocity field, or, better yet, integrate the gas streamlines, and identify the part of the gas that does not move away from the planet, hence ``bound'' to it. This type of approach finds a CPD size of about $0.1 r_{\rm B}$ in the midplane for sub-thermal planets \citep{fung19cpd,2021ApJ...915..113B}, and about $0.5 r_{\rm H}$ for super-thermal planets \citep{machida08}. However, this approach may not be ideal when considering that the gas in the CPD can ''recycle'' \citep{ormel15}.

As mentioned, gas typically flows vertically toward the planet near the poles and away from the planet near the midplane. In some isothermal simulations, this is found to happen everywhere \citep{ormel15,kurokawa18}, and no gas is formally bound to the planet. The gas near the planet is then said to be recycling, where gas co-orbiting with the planet flows into the CPD and back out within a finite time. This recycling time, which is defined as the enclosed mass within a given radius around the planet, such as the Bondi radius, divided by the outward mass flux at that radius, has a steep dependence on planet mass, and can range from one orbital period of the planet to many thousands of periods \citep{ormel15,fung15cpd}. It is unclear how to define a CPD size in these cases. The concept of recycling remains important for non-isothermal CPDs, which we will discuss in the following section.

Finally, we reflect on the interpretation of these isothermal results. Since CPDs are generally assumed to be hot, dense environments, the isothermal assumption may be thought of as the fast cooling scenario, where heat from adiabatic compression is radiated away instantly, or at least within a time much shorter than the dynamical time of the gas. It is unclear whether such a limit can be reached in a realistic setting. Cooling is regulated by dust concentration in the gas---too little dust and the gas cannot cool, but too much dust will also trap heat. Moreover, isothermality does not only imply fast cooling, but also fast heating. Around an isothermal CPD, the gas is effectively receiving instant heating from an unknown source whenever it expands. Such a mechanism has no clear justification, and it is unclear what might change if the gas cannot heat up in such a manner.

An alternative way to interpret isothermal simulations of CPDs is that they may resemble the end points of CPD evolution. The CPD may begin as a hot system, but after a sufficiently long time, its temperature should equilibrate with the surrounding circumstellar disk, assuming it has not yet dissipated. In this interpretation, the thermal history of the CPD, i.e. how it cools down over time, may play a role in the CPD's final state, but is clearly not captured in isothermal simulations. Despite that, some features in isothermal CPDs do appear consistent with observations. The giant planets in our solar system all host extensive satellite systems likely born out of rotationally-supported CPDs. So far, rotationally-supported CPDs have only been found in isothermal simulations, and the $r_{\rm c}$ measured are consistent with the sizes of the satellite systems around our gas giants \citep{fung19cpd}. This suggests that moon and satellite formation may have occurred in nearly isothermal CPDs.

\subsubsection{Non-Isothermal CPDs}

Figure~\ref{fig:CPD} demonstrates how sensitive the CPD structure is to the equation of state of the gas. Merely switching from an isothermal one to an adiabatic one produces a drastically different density structure around the planet. In pursuit of realistic thermodynamics, we must consider a suite of model parameters, including the adiabatic index, viscous heating of gas, dust-to-gas ratio, dust opacity $\kappa$, and accretion luminosity of the planetary core $L_{\rm acc}$. In non-isothermal simulations, the dynamically varying disk temperature makes it difficult to define $q_{\rm th}$. Roughly, we can estimate $h$ and $q_{\rm th}$ using the temperature of the disk at the planet's location, before the planet is introduced, given by $h\sim c_{\rm s}/v_{\rm k}$ where $c_{\rm s} = \sqrt{P/\rho}$ is the sound speed if the gas were isothermal; though, one should bear in mind that the temperature of the disk after the planets are introduced can become very different, and is dependent on radiative transport parameters that are generally not the same between different simulations.

Simulations of non-isothermal CPDs have overwhelmingly resulted in CPDs that are supported by gas pressure with weak or no rotation at all \citep{ayliffe09,dangelo13,szulagyi16,lambrechts17,cimerman17}. These CPDs are not disk-like, but spherically symmetric envelopes, making the term circumplanetary ``disks'' rather inaccurate. In some exceptions, the CPDs can become more disk-like. \cite{szulagyi16} found that, for $q_{\rm th}=8$, if the temperature near the planet is fixed to a value $\lesssim 2000$ K, one can recover rotationally-supported disks. \cite{szulagyi17} showed that the disk morphology becomes flattened with shock surfaces forming above the and below the CPD when $q_{\rm th}\gtrsim24$. Then, in a lower mass range where $q_{\rm th}=0.1\sim0.5$, \cite{lambrechts17} found that when the product $\kappa L_{\rm acc}$ exceeds a certain threshold value, heat transfer in the CPD becomes advection-dominated and the morphology of the disk becomes flattened.

One can measure the size of these envelopes using kinematics same as we do with isothermal disks. \cite{dangelo13} found that the bound envelopes are about $0.3\sim0.7 r_{\rm B}$ for $q_{\rm th}\approx0.05\sim0.2$, where the envelopes take up a smaller fraction of $r_{\rm B}$ with increasing $q_{\rm th}$. \cite{lambrechts17} measured $\sim0.2 r_{\rm B}$ for their $q_{\rm th}\sim0.56$ planets. The results of \cite{dangelo13} and \cite{lambrechts17} appear in line with each other. \cite{cimerman17} did not identify strictly bound envelopes, but instead defined the envelope as the region that can cool over time. They measured sizes of $\sim0.2 r_{\rm B}$ when $q_{\rm th}=0.38-0.75$, and $\sim0.07 r_{\rm B}$ when $q_{\rm th}=1.9$. The more simplified adiabatic simulations by \cite{fung19cpd} similarly found envelope sizes of $\sim 0.2 r_{\rm B}$ for $q_{\rm th}\lesssim1$. Overall, $0.2 r_{\rm B}$ is the common answer for $q_{\rm th}\sim0.1-1$ for both radiative and adiabatic models, which suggests that CPDs begin their evolution under roughly adiabatic conditions. For super-thermal planets, \cite{lambrechts19} reported bound envelope sizes of $\sim 0.3 r_{\rm H}$. Instead of treating radiative transport, \cite{kurokawa18} modeled cooling (and heating) using a simple thermal relaxation treatment. They found that the size of the envelope depends on the thermal relaxation time. When the cooling time is $0.01 \Omega_{\rm p}^{-1}$, it can become as large as $r_{\rm B}$, larger than both isothermal and adiabatic CPDs. The envelope size as a function of the cooling rate is therefore non-monotonic.

The gas accretion rate is another important measurement one can make in radiative simulations of CPDs. This is a frontier of research and a unified picture has not yet emerged. For instance, \cite{lambrechts19} reported that Jupiter-mass planets can form from $15$ Earth-mass cores within a fraction of the disk lifetime, and showed that their measurements were in qualitative agreement with 1D gas accretion models, such as those by \cite{Piso14} and \cite{Lee14}. Their models did find bound, pressure-supported envelopes of order the size of $r_{\rm B}$, albeit smaller than commonly assumed in 1D models, which supports the idea that envelopes are effectively 1D. In contrast, \cite{Moldenhauer21} reported that 1D models and 3D simulations produce drastically different outcome, where the 3D, recycling flow of the gas prevents the planet's envelope from cooling. In their models, they found that recycling occurs throughout the entire envelope, all the way down to the core. In this scenario, none of the gas is truly bound to the planet, and it always has a finite time to cool before being recycled back to the circumstellar disk. This can eventually stunt the growth of the envelope and prevent the planet from ever developing into a gas giant. 
%JF: added text
While \cite{lambrechts19} and \cite{Moldenhauer21} both simulated sub-thermal planets ($q_{\rm th}$=0.7 and 0.18, respectively), different assumptions were made about other local conditions, including disk temperature and opacity. These factors will like be critical in understanding how these differing reports may be united into a single, self-consistent model of 3D gas accretion.

%SJP: added ref to Gressel (2013)
Finally, it is worth mentioning that other physical effects, such as magnetic fields, can be important shaping CPDs. Global MHD simulations with adaptive mesh refinement were presented in \cite{2013ApJ...779...59G}, who found a CPD whose structure displays high levels of variability, and the generation of helical magnetic fields that launch magnetocentrifugally driven outflows.

\section{Tools and Techniques}
\label{sec:tools}
The progress that has been made in this field has been primarily enabled by hydrodynamic simulations. This section discusses briefly the basic challenges of this type of calculation \citep{2015ASPC..498..216M,2018JPhCS1031a2007M}, the state of the art, and points to the current community codes. 

\subsection{Challenges}

Hydrodynamic simulations of planet-disk interactions share many challenges with hydrodynamic problems in other fields. One of these is the problem of a wide range of important scales that need to be resolved. Global disk evolution occurs on scales of $\sim 100$ AU. For a planet located at $1$ AU, the length of an orbit, over which libration in the horseshoe region occurs, is $2\pi \times$ AU. The pressure scale height in the disk at this location is $H\sim 0.1$ AU. The width of the horseshoe region for a Earth-like planet is $\ll H$ \citep[see e.g.][]{2009MNRAS.394.2297P}. The plume responsible for the thermal torque depends on the thermal diffusivity but is usually much smaller still \citep{2017MNRAS.472.4204M}. Finally, the physical radius of an Earth-like planet is $\sim 10^{-3} H$. It is likely that all of these scales are talking to each other. For example, global disk evolution regulates the inflow of solids, which in the end are responsible for the plume governing the thermal torque. Different methods have different ways of trying to meet this challenge: Adaptive Mesh Refinement for grid-based codes, while Smoothed Particles dynamics naturally yields high resolution in regions of high density.   

A second challenge arises due to the range of important time scales. The longest important time scale is the disk life time, i.e. millions of years, which is computationally unfeasible. Planet-disk interactions simulations need to be run usually until a quasi-steady state is reached. For gap-opening planets, this can be many thousands of orbits \citep[e.g.][]{2021arXiv210505277D}. Every orbit contains $\sim 1000$ time steps, depending on the numerical resolution \citep{2015ASPC..498..216M}. Each simulation then takes many millions of time steps, quickly making 3D computations very expensive. As a rough guide, a grid-based simulation, including only basic hydrodynamics, consisting of $\sim 10^9$ computational cells \citep[i.e. $\sim 1000$ cells in each direction, see][]{2020arXiv200211161M} will take $\sim 1000$ seconds per time step on a modern CPU. Each simulation then takes $\sim 10^6$ CPU hours. Additional physics in the form of radiative cooling, magnetic fields, etc, will further increase the computational cost. 

Since horseshoe drag relies on the material conservation of vortensity and entropy\index{Conservation laws} \citep{1991LPI....22.1463W, 2008ApJ...672.1054B,2008A&A...485..877P,2009MNRAS.394.2283P, 2009ApJ...703..845C,2009ApJ...703..857M}, it is important that numerical schemes adhere to these as much as possible. Traditionally, staggered mesh codes, have an advantage here over those built on Riemann solvers \citep{2015ASPC..498..216M}. One possible explanation for this is that most Riemann solvers are built using directional splitting and use 1D Riemann solvers, while for example vorticity is a multidimensional quantity. A truly multidimensional Riemann solver can substantially improve for example the lifetime of vortices in disks or MHD structures  \citep[e.g.][]{BALSARA20101970,2017MNRAS.469.4306P}. SPH in principle has excellent conservation properties for quantities that can be specified on a per-particle basis (e.g. angular momentum, entropy), and under certain conditions for potential vorticity \citep{frank03}. However, since SPH simulations of protoplanetary disks usually result in an artificial viscosity equivalent to $\alpha \sim 10^{-2}$ \citep{2018PASA...35...31P}, conservation of potential vorticity in horseshoe dynamics is washed out by the action of viscosity. 

Another potential issue arises in particular for hydrostatic equilibria, which could be the planet atmosphere or the vertical disk profile. The numerical scheme should preserve such stable states, meaning that the scheme would be 'well-balanced'. This is traditionally a problem for fractional step methods as implemented in many Riemann solvers, but solutions do exist  \citep[e.g.][]{1995A&AS..110..587E, BaleLevMitRoss02,2006A&A...450.1203P}.

Shocks play an important role in particular for high-mass planets, capable of opening up a gap \citep{1996ApJS..105..181K}. However even for low mass planets in inviscid disks shocks do appear \citep{2001ApJ...552..793G}. It is therefore crucial that numerical methods correctly treat shocks. Here Riemann solvers have an obvious advantage over other methods, since the correct treatment of shocks forms an integral part of the solution method. This allowed \cite{2011ApJ...741...57D} to obtain the correct evolution of a density wave excited by a low-mass planet towards a shock using ATHENA \citep{2008ApJS..178..137S}. 

As protoplanetary disk models become more sophisticated, so does the amount of physics that should be part of planet-disk interaction simulations. The field has moved from 2D and 3D, locally isothermal hydrodynamical models \citep{2002A&A...385..647D,2003ApJ...586..540D}, to ideal MHD simulations \citep[e.g.][]{2004MNRAS.350..849N}, to radiation-hydrodynamical models \citep[e.g.][]{2008A&A...478..245P}, to non-ideal MHD models \citep{2017MNRAS.472.1565M}. In addition, multiple phases are often included \citep[i.e. gas and solids, see e.g.][]{2004A&A...425L...9P,2018ApJ...855L..28B}. Needless to say, the inclusion of all these important effects drives up the computational costs. 

Finally, the paradigm shift away from viscous, turbulent accretion disks towards laminar, wind-driven accretion disks brings a relatively new challenge. A disk where diffusion is negligible has a very long memory. This means that initial conditions will matter a lot more compared to previous, viscous disk models. Planets will remember where they have been \citep[see e.g.][]{2014MNRAS.444.2031P}, which makes realistic initial conditions more important than ever.

\subsection{Community codes}

Below, we point to community codes that have been used recently to study planet-disk interactions. Some of these have been around since the comparison problem of \cite{2006MNRAS.370..529D}. We mainly focus on codes that are freely available for download. The list is not exhaustive, but should give the reader an idea of the kinds of codes that are currently used. The feature lists for each code is incomplete as well: many codes have a modular structure and new modules are added on a regular basis.  

\subsubsection{Staggered Mesh Codes}

We start off by describing the family of grid-based methods that use a staggered mesh. These were termed 'upwind methods' in \cite{2006MNRAS.370..529D}. These share their core hydrodynamical algorithm with ZEUS \citep{1992ApJS...80..753S} and are based on finite differences.  

{\bf FARGO3D:} The original version of FARGO \citep{2000A&AS..141..165M} was specifically aimed at planet-disk interactions, and its successor FARGO3D \citep{2016ApJS..223...11B} has inherited the features that make it straightforward to set up a planet-disk interaction problem. FARGO3D can do MHD and can run on GPUs. It has a module to include multiple dust species \citep{2019ApJS..241...25B}. Recent applications to planet disk interactions include \citet{2020MNRAS.493.4382M} and \citet{2021MNRAS.500.1621D}. 

Other well-known staggered mesh codes include the original NIRVANA code \citep{1997CoPhC.101...54Z}, which was never publicly released, but was part of the 2006 comparison problem in several incarnations \citep{2006MNRAS.370..529D}, the FARGOCA version of FARGO maintained in Nice \citep{2014MNRAS.440..683L}, and the code used in \cite{2018A&A...618A...7V}. 

\subsubsection{Riemann solver Codes}

Many of the current community codes have Riemann solvers at their core. We list them here in alphabetical order.

{\bf ATHENA++:} This is the successor of the Athena MHD code \citep{2008ApJS..178..137S}, ported to C++ and with new features added \citep{2020ApJS..249....4S}. It has a Riemann solver at its core, and is fourth order accurate in space and time in regions of smooth flow \citep{2018JCoPh.375.1365F}. The code was first used to simulate planet-disk interactions by \citet{dong15spiralarm} and \citet{zhu15densitywaves}. More recent publications include \citet{2021ApJ...915..113B} and \citet{2020A&A...643A..21K}. 

{\bf NIRVANA:} The public version of NIRVANA uses Riemann solvers \citep{2004JCoPh.196..393Z}, and has support for non-ideal MHD, adaptive mesh refinement and self-gravity. It has been used recently in \cite{2017MNRAS.472.1565M}.

{\bf PLUTO:} The finite volume code PLUTO \citep{2007ApJS..170..228M} uses a Riemann solver at its core, and has support for MHD and solid particles. A version that runs on GPUs was presented in \cite{2017A&A...604A.102T}. Examples of recent work on planet disk interactions include \citet{2018MNRAS.478.2737C} and \citet{2021A&A...645A..68P}.

Other, not publicly available codes using Riemann solvers include RODEO \citep{2006A&A...450.1203P} and PEnGUIn \citep{2014ApJ...782...88F}.

\subsubsection{High-order Finite Difference Codes}

This class of methods achieves high accuracy by taking high-order finite differences. The resulting scheme needs to be stabilized by artificial viscous terms.

{\bf PENCIL:} This is a high-order finite difference MHD code \citep{2021JOSS....6.2807P} and has support for solid particles in addition to gas \citep{2007ApJ...662..613Y}. For recent applications on planet-disk interaction see \citet{2016ApJ...817..102L} and \citet{2020MNRAS.491.4702Y}.  

\subsubsection{Smoothed Particle Hydrodynamics}

SPH methods are popular for disk simulations, since they naturally allow for complex geometries such as warps, and because of their oct-tree data structure, the inclusion of self-gravity is straightforward. 

{\bf PHANTOM:} This Smoothed Particle Hydrodynamics (SPH) code \citep{2018PASA...35...31P} has support for MHD and dust particles \citep{2020MNRAS.499.3806M} and is widely used to simulate accretion disks. Recent publications on planet disk interactions include \citet{2021MNRAS.504..888B} and \citet{2020MNRAS.499.2015T}.

{\bf GADGET:} The publicly available version GADGET-2 \citep{2005MNRAS.364.1105S} has support for self-gravity. For a recent application of GADGET to disk-planet interaction see \cite{2019MNRAS.489.5187H}.

{\bf SEREN:} This SPH code was developed for star and planet formation simulations \citep{2011ascl.soft02010H,2011A&A...529A..27H}. It supports self-gravity and cooling through radiation. A recent application to disk-planet interactions is \cite{2018MNRAS.477.3110S}.

For other, not publicly available SPH implementations see for example \cite{2019MNRAS.486.4398F}.

\subsubsection{Other methods}

{\bf DISCO:} The moving-mesh MHD code DISCO \citep{2016ApJS..226....2D} employs a moving-mesh approach utilizing a dynamic cylindrical mesh that can shear azimuthally to follow the orbital motion of the gas, in combination with a Riemann solver. It is specifically designed to deal with the shearing motion present in astrophysical disks. Recent publications on planet-disk interactions include \citet{2020ApJ...889...16D} and \citet{2015ApJ...812...94D}.

High-order discontinuous Galerkin (DG) methods \citep[e.g.][]{1998JCoPh.141..199C} are slowly finding their way into astrophysical fluid dynamics. An implementation for planet-disk interactions was presented in \cite{2018MNRAS.478.1855V}. While the use of DG methods in planet-disk interactions is still in the exploratory phase, we have included them here as because of their combination of high-order accuracy with shock-capturing capabilities, they may well represent the future of this type of calculations.

Finally, it is worth mentioning the moving mesh code AREPO \citep{2010MNRAS.401..791S}, which has been used on the disk-planet problem in \cite{2014MNRAS.445.3475M}, and the meshless code GIZMO \citep{2015MNRAS.450...53H}, which was part of the comparison paper \cite{2019MNRAS.486.4398F}.

%Resolution close to the planet.
%Full azimuthal extent (libration).
%Large number of time steps per orbit.
%Large number of orbits. 
%Vorticity/entropy advection.
%Shocks
%Hydrostatic equilibrium.
%Physics: non-ideal MHD.
%Small time steps: FARGO, super time stepping.
%Low viscosity; long memory.

%For example, in planet-disk interactions orbital advection is commonly used in the form of a remap step
%\citep{2000A&AS..141..165M,2016ApJS..223...11B},
%or a  restricted moving mesh \citep{2016ApJS..226....2D}.

%\subsection{Fixed mesh codes {\color{red} (Sijme-Jan)}}
%FARGO3D, Athena++, Pluto, DG. 

%\subsection{Moving mesh codes {\color{red} (Paul?)}}
%DISCO, AREPO (ever been used?).

%\subsection{Meshless methods {\color{red} (?)}}
%SPH, Gizmo (?).

%acknowledge PLUTO is a a popular Godunov code for planet-disk
%
%exploration of DG methods \citet{2018MNRAS.478.1855V}
%
%limited study of full non-ideal MHD
%some treatment of wind in a reduced model
%
%Use of 1D extended boundaries ex fargo1d2d and \citet{2016ApJ...826...13B}.

\section{Modeling Observational Signatures\index{Disk!Imaging}\index{Disks!Scattered light}}
\label{sec:obs}

In this section we introduce recent work on modeling observational signatures of disk-planet interactions. Advancements in theory and numerical simulations in the last decade play a crucial part in the emergence and development of this new field. More detailed discussions can be found in the chapter led by Bae et al. We focus on dust observations (ONIR scattered light and mm/cm continuum emission). Planet-induced gas kinematics signatures are discussed in more details in the chapter led by Pinte et al.

Observational signatures of planet-induced structures are generally modeled in a two-step process \citep[e.g.,][]{dong15gap}. First, hydrodynamics simulations are carried out to calculate the spatial distribution and kinematics of gas perturbed by one or more planets. Dust particles of various sizes may be added into the simulations to obtain the spatial distribution of dust as they dynamically interact with the gas. In the second step, the resulting hydro disk models are fed into radiative transfer simulators, which produce synthetic images of various kinds. The hydro models may be intrinsically 3D, which provide volume density and velocity distributions, or 2D, in which case they are often ``puffed up'' in the vertical direction to obtain 3D structures. The resulting synthetic images may be post-processed to achieve realistic angular resolutions and sensitivities, to facilitate direct and quantitative comparisons with observations.

\subsection{Generic modeling of planet-induced structures in observations}

Planets on stationary, coplanar, and circular orbits are expected to produce disk structures, including spiral arms, gaps, vortices, and circumplanetary disks. Inclination or eccentricity in the planets' orbits and planet migration can affect the appearance of these features, or even produce additional ones.

\subsubsection{Spirals\index{Disks!Spiral waves}}

One planet can drive multiple spiral arms both inside and outside its orbit, as different modes in the resonances\index{Disks!Resonances} excited by the planet constructively interfere \citep{bae18theory, miranda19theory}. The number and strength of the spirals depend on the planet mass (\mplanet) as well as disk properties such as aspect ratio ($h/r$), and thermal relaxation rate \citep{bae18simulation, zhang20}. At tens of AU, spiral arms excited by planets more massive than Saturn may be traceable in ONIR scattered light using current generation of instruments \citep{dong17spiralarm} at a variety of viewing angles \citep{dong16armviewing}. As the planet mass increases, the strength of spirals in scattered light and the separation between the primary and secondary spirals increase \citep{fung15, dong17spiralarm, bae18simulation}; such quantitative correlations can be exploited to dynamically constrain the planet masses. Specifically, a planet with $M_{\rm p}\sim$ a few $\times M_\star (h/r)^3$ (usually supra-Jovian) can drive a pair of prominent arms in a near symmetric configuration inside its orbit \citep{dong15spiralarm}. 

Spirals may also be detected as temperature perturbations in disks with non-isothermal EOS as gas been compressed and decompressed crossing spirals \citep{bae21, muley21}. Inside spiral arms, significant vertical motion is present and may be traceable \citep{bae21}. Because of this, modeling observational signatures of spirals generally requires 3D disk-planet interaction simulations; puffing up 2D hydro models assuming hydrostatic equilibrium would fail to capture the vertical motion and misrepresent the gas density structure, producing scattered light signatures that are too weak \citep{dong17spiralarm}.

\subsubsection{Gaps}

Planet-opened gaps appear as annular structures with low intensities in observations. Under a modest disk viscosity ($\alpha\gtrsim10^{-3}$), usually only one gap at planet's orbit forms. The radial profile of gas surface density, as well as the depth and width, of such a gap may be parametrized using \mplanet, $h/r$, and $\alpha$ \citep{2014ApJ...782...88F,duffell15gap,2020ApJ...889...16D,  kanagawa15,  kanagawa16width, kanagawa17deepgaps}. In ONIR scattered light, the contrast of a single gap with the adjacent disk can be analytically expressed using these parameters and the angular resolution of the observations \citep{dong17gap}. Gaps are also detectable in mm/cm dust continuum observations that probe the spatial distribution of mm/cm-sized dust. In those observations, the adjacent dust rings may become broad due to dust feedback \citep{kanagawa18dustfeedback}, and have finite vertical thickness due to planet-driven meridional turbulence \citep{bi21}. Aerodynamical coupling between the gas and dust leads to dust concentrations at gas pressure peaks, resulting in higher contrasts in dust emission observations than in scattered light \citep{rosotti16}. Models of the observational signatures of gaps are generally insensitive to 3D dynamics \citep{fung16, dong17gap}, but 3D simulations are still required to capture the meridional motion.

In theory, the dissipation of each spiral arm excited by a planet can lead to the opening of an individual gap, thus one planet may open a few gaps as it drives a few spirals \citep{dong17doublegap, bae17}. Such signatures are most prominent when there is no mechanisms to dissipate spiral arms other than their intrinsic nonlinearity, e.g., in disks with low viscosity ($\alpha\lesssim10^{-4}$; \citealt{2001ApJ...552..793G}). The density waves and the separation between these gaps are regulated by \mplanet, $h/r$, and thermal transport \citep{dong18doublegap, zhang18, miranda19adiabatic, miranda20, miranda20inplane, ziampras20}. In addition, if the planet is migrating, the spacing and morphology of the gaps can be further modified \citep{meru19, nazari19, perez19hd169142, weber20}, and more gaps may be produced if the planet migrates in a non-steady way \citep{wafflardfernandez20}. Finally, a low mass planet may open gaps in dust of certain sizes even when it does not open prominent gas gaps \citep{dipierro17}.

\subsubsection{Vortices}

Vortices can be excited by the Rossby wave instability at the edge of planet-opened gaps. They trap dust particles and can appear as emission clumps in thermal continuum observations \citep{zhu14stone}. The location of the dust concentration depends on the dust size, thus observations at different wavelengths may detect azimuthal variations in the vortex location \citep{mittal15, baruteau16}. In addition, vortices produce spiral arms, though they are typically too weak to be detectable \citep{huang19}.

When the dust-to-gas mass ratio reaches order unity at the gap edges, dust feedback enhances the sharpness of the edges, and leads to instabilities and the formation of tiny vortices with highly concentrated dust in continuum observations \citep{huang20, 2020MNRAS.497.2425H}.

\subsubsection{Signatures of planets on eccentric and/or inclined orbits}

A massive planet on an inclined orbit may break the inner disk out from the outer disk, and cause the two parts to precess at different timescales \citep{zhu19}. Consequently the two parts of the disk may become warped or misaligned. The inner disk may cast shadows on the outer disk, and the radially varying disk inclination may be visible in gas kinematic observations,
in a way similar to the effect of a misaligned binary \citep{facchini18}. A giant planet on an eccentric orbit can carve out a gap as wide as its orbit, much wider than its circular counterpart \citep{muley19}. Such a gap can also appear much broader in C$^{18}$O than in $^{13}$CO observations \citep{baruteau21}.

\subsubsection{Signatures of planets in disk kinematics}

Planet-induced kinematic signatures have been modeled using hydrodynamics simulations and synthetic gas observations. 
\citet{perez15cpd} showed that the rotation of a CPD around a giant planet produces a unique signature in disk channel maps well separated from the circumstellar disk. 
\citet{pinte18} showed that a planet may create localized ``kinks'' in the isovelocity maps of its host disk. 
\citet{teague18}
quantified the radial modulation in the circumstellar disk rotation
due to the pressure gradients at gap edges. 
\citet{perez18} investigated gas kinematics
caused by pressure gradients at gaps, spiral wakes, and vortices.
\citet{dong19kinematics} and \citet{teague19} showed that a planet opening a deep gas gap can drive transonic vertical motions inside the gap potentially detectable in disk Moment 2 maps. More discussions on these topics can be found in the chapter led by Pinte et al. and in the recent review article by \citet{armitage20}.

\subsection{Planet-disk interaction as the origin of observed disk structures}\label{sec:planetdisk}

Hundreds of protoplanetary disks have been imaged by the latest generation of instruments. In most of them, structures that resemble planet-induced have been found. Observations are now being quantitatively compared with disk-planet interaction models to infer the presence of planets forming in disks, and to constrain their properties, such as mass and orbit \citep{bae18, zhang18, lodato19}. These planets are extremely difficult to find directly \citep[e.g.,][]{keppler18}; yet they are the keystones in testing planet formation theories, as they provide direct constraints on the location, timescale, and local environment of planet formation.
 
\begin{figure}[h]
 \epsscale{1.0}
 \plotone{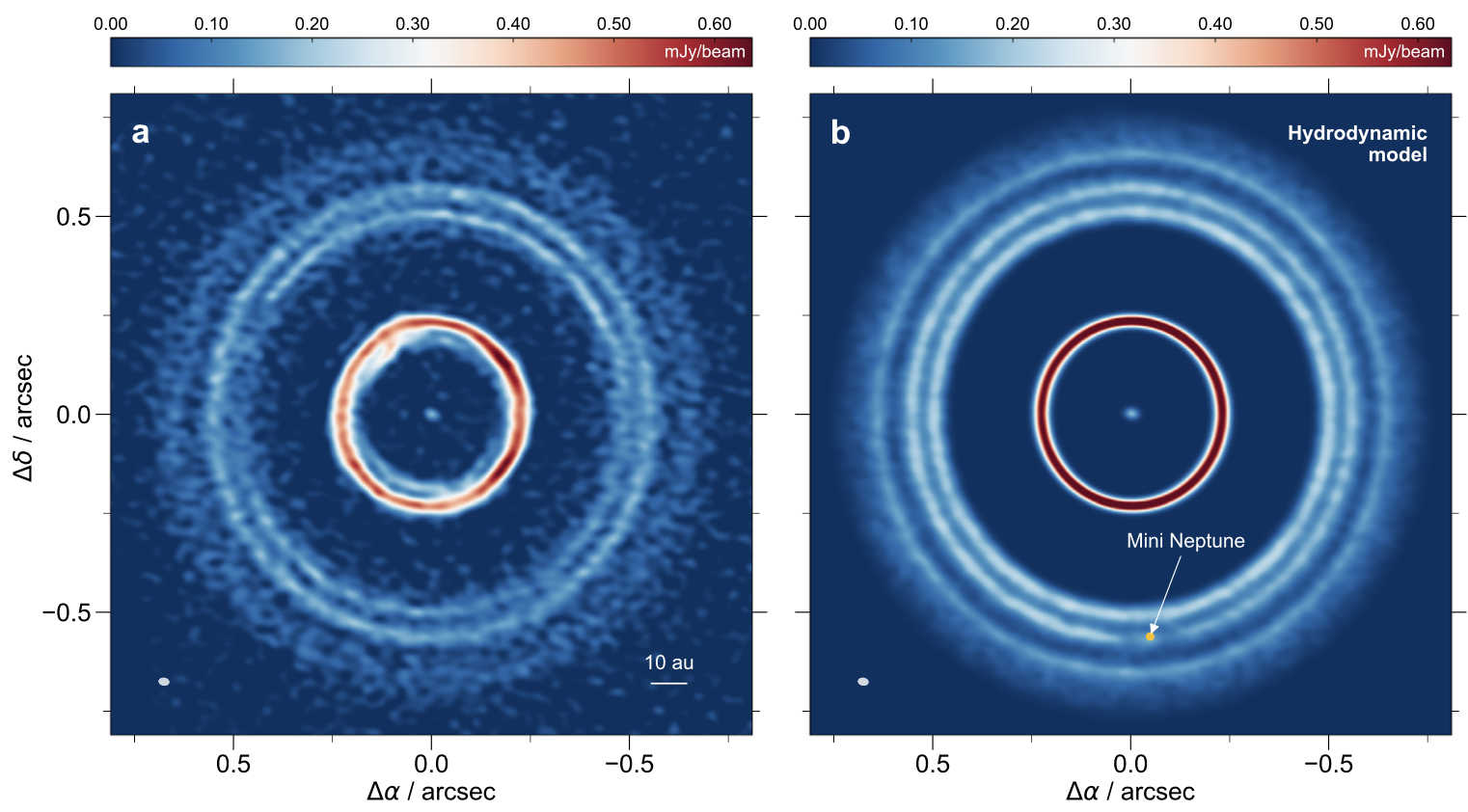}
 \caption{\small Left: ALMA 1.3 mm image of the HD 169142 disk. Right: Synthetic observations of a disk with a migrating mini-Neptune planet of 10 $M_\oplus$ \citep{perez19hd169142}. See section \ref{sec:planetdisk} for details.}
 \label{fig:hd169142}
\end{figure}

There have been numerous models where disk features are explained with planet-disk interactions.
Good quantitative agreement between theory and observation in many of them supports the planetary origin of those features.
Below we review some representative examples focusing on structures seen in images. Disk-planet interaction models have also been tested using other observational diagnostics, e.g., the spectral energy distribution of the disk \citep[e.g., in the case of CI Tau,][]{muley21citau}\index[obj]{CI Tau}.
\begin{enumerate}
\item HL Tau\index[obj]{HL Tau}. As the first protoplanetary disk targeted by ALMA\index{Atacama Large Millimeter Array} with its long baseline configuration \citep{brogan15}, the HL Tau disk attracted immediate attention when several gaps at tens of AU were revealed. Several groups attempted to explain its gaps using planets. Assuming a modest viscosity of $\alpha\sim10^{-3}$, \citet{dong15gap}, \citet{dipierro15hltau} and \citet{jin16} reached the same conclusion that the three main gaps can be explained as opened by three Saturn mass planets at around 13, 33, and 70 AU (Figure~\ref{fig:hltau}).

\begin{figure}[h]
 \epsscale{1.0}
 \plotone{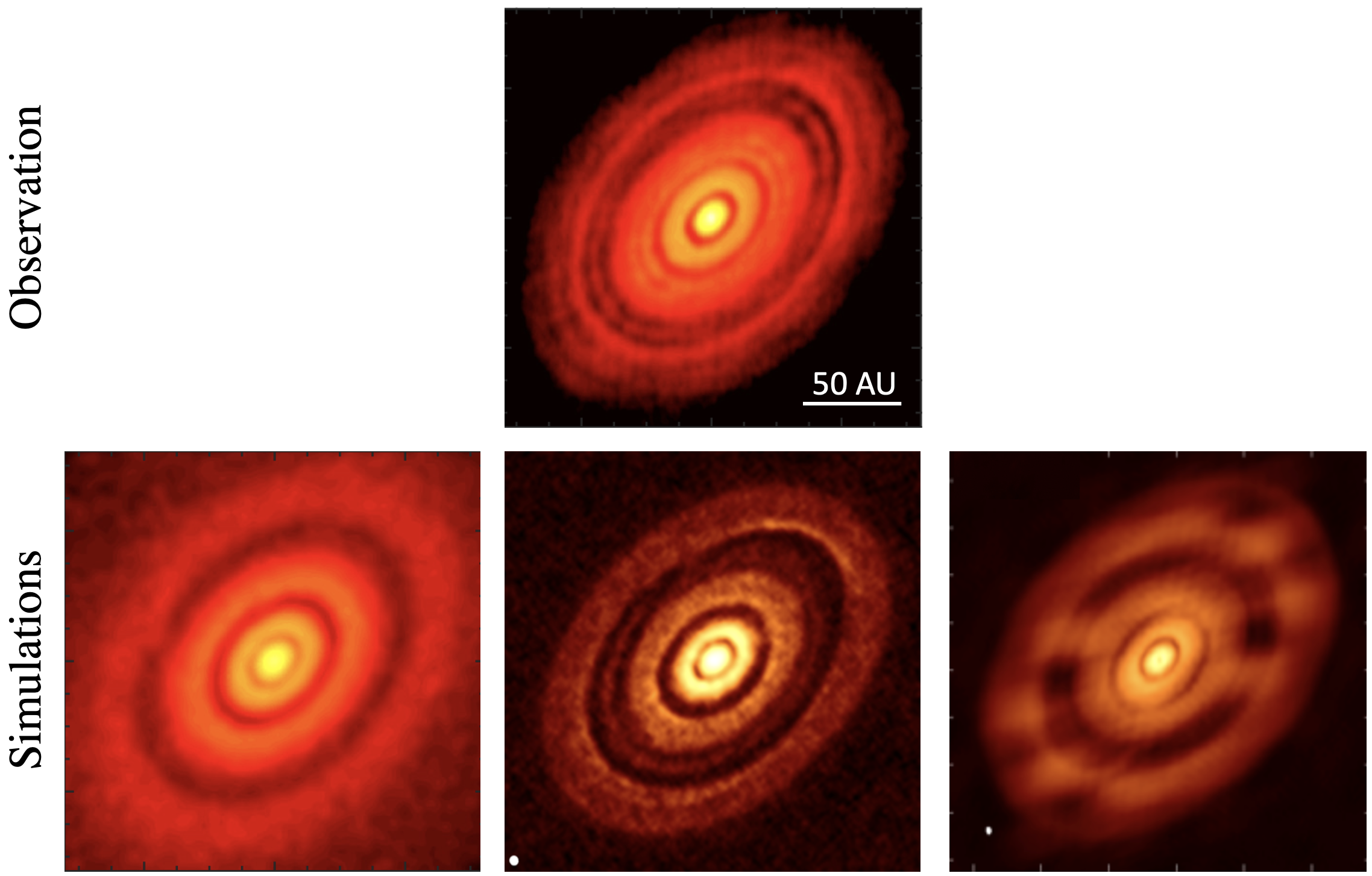}
 \caption{\small Top: ALMA\index{Atacama Large Millimeter Array} observations of the HL Tau disk, showing multiple rings and gaps in dust continuum emission at mm wavelengths \citep{brogan15}. Bottom: Simulations by three independent teams that show the three main gaps in the disk may be explained as opened by three Saturn mass planets, if the disk has a modest viscosity $\alpha\sim10^{-3}$ \citep{dong15gap, dipierro15hltau, jin16}. See section \ref{sec:planetdisk} for details.}
 \label{fig:hltau}
\end{figure}

\item MWC 758 \index[obj]{MWC 758}. The MWC 758 disk was one of the first to be discovered to harbor a pair of near-symmetric spiral arms in scattered light \citep{grady13, benisty15}. \citet{dong15spiralarm} and \citet{zhu15densitywaves} showed that a pair of such arms can be excited by a super-thermal mass planet (usually several Jupiter masses or higher) on the outside at $\gtrsim$100 AU. Later, ALMA mm continuum observations showed that the disk harbors a cavity $\sim40$ AU in radius, and in the outer disk there are two emission clumps at $\sim50$ and 80 AU \citep{boehler18, dong18mwc758}. The northern clump has a counterpart in cm emission revealed by the VLA observations, but not the southern clump \citep{casassus19}. \citet{baruteau19} explained these additional features by involving a second planet $\sim2M_{\rm J}$ in mass at 35 AU. Together with the outer planet, they each open a deep gap, and trigger the formation of a vortex trapping mm-sized dust at the gap edge (inner edge for the outer gap and outer edge for the inner gap). The southern clump is in the process of decaying, thus it does not trap cm-sized dust well. MWC 758 is an excellent example of using disk-planet interaction models to explain disk structures in multi-wavelength observations (Figure~\ref{fig:mwc758}).

\begin{figure}[h]
 \epsscale{1.0}
 \plotone{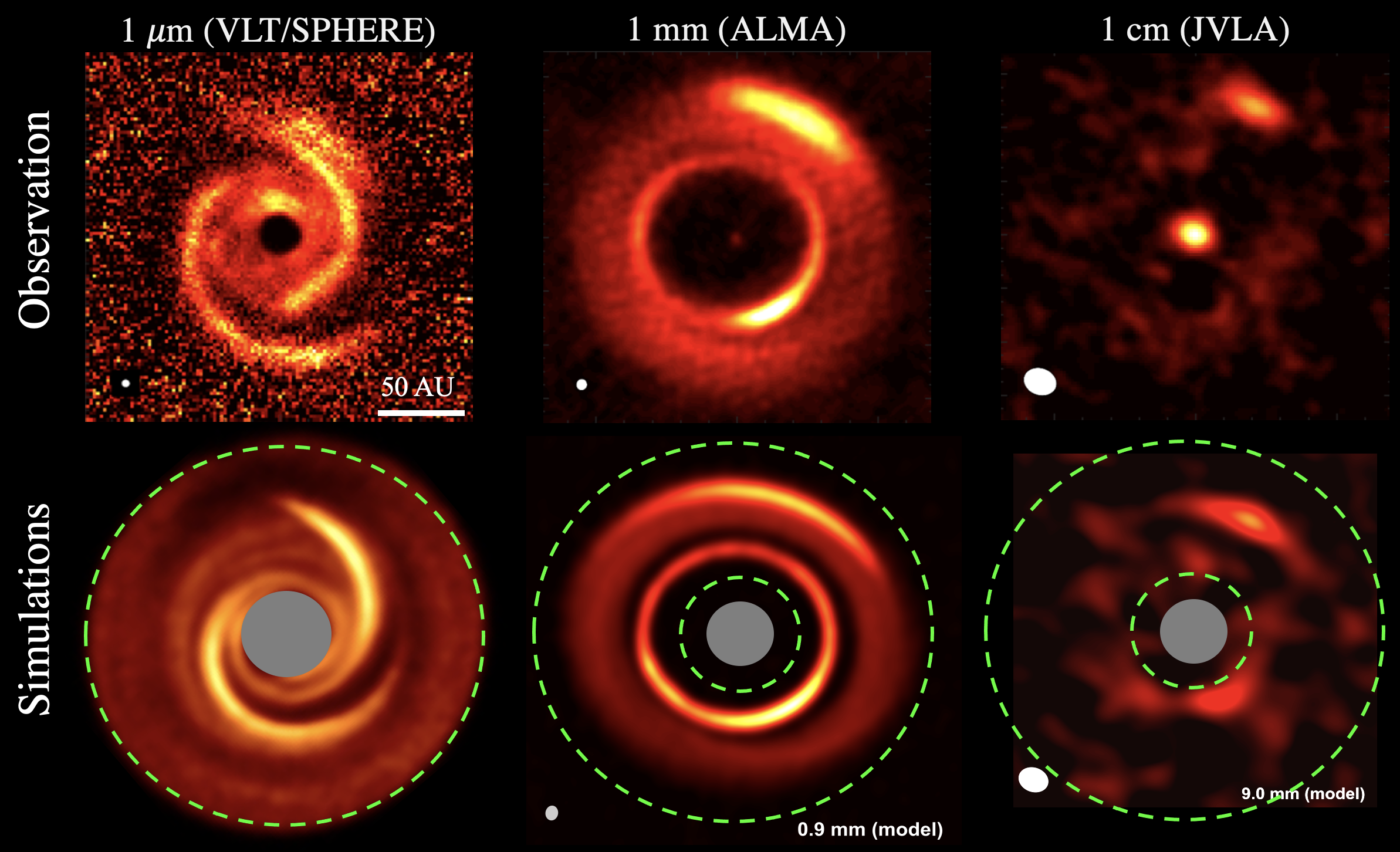}
 \caption{\small Top row: Three observations of the MWC 758 disk at different wavelengths on the same scale. From left to right: a scattered light image at $\sim1\micron$ \citep{benisty15}, an ALMA continuum emission map at $\sim1$mm \citep{dong18mwc758}, and a JVLA continuum emission map at $\sim1$cm \citep{casassus19}. Bottom row: Three synthetic observations of the same kind as the panels above \citep{dong15spiralarm, baruteau19}. The orbits of the planets in the models are traced out by the green dashed circles. There is one giant planet in the model on the left, and there are two giant planets in the model shown in the middle and right panels. The models match the observations reasonably well. See section \ref{sec:planetdisk} for details.
}
 \label{fig:mwc758}
\end{figure}

\item HD 169142 \index[obj]{HD 169142}. \citet{perez19hd169142} discovered a double gap and triple ring structure in a compact configuration at around 70 AU in the HD 169142 disk using ALMA mm continuum observations, and showed that the structure can be produced by an inwardly migrating $\sim$10 Earth mass mini-Neptune at $\sim68$ AU in a disk with low viscosity. This is one of first cases in which migration is required in order to explain observed disk structures, and the predicted planet mass is among the lowest (Figure~\ref{fig:hd169142}).
\end{enumerate}

While in the above examples, disk-planet interaction models reproduced real observations, considerable uncertainties still exist.
First of all, the properties and even the number of planets needed to explain certain structures depend on disk properties, some of which remain poorly constrained. For example, in the case of HL Tau \citet{bae17} showed that most observed gaps can be explained by a single $\sim$Neptune mass planet if the disk viscosity if lower than $\alpha=10^{-4}$. A similar example has been shown in the case of AS 209\index[obj]{AS 209}, in which 5 gaps may be produced by one planet \citep{zhang18}. Further discussions on this topic can be found in the chapter led by Bae et al..
Secondly, alternative theories that do not involve planets may also reproduce observed structures. Whether these mechanisms are operating in a significant fraction of disks is still under debate. For example, in the case of MWC 758, the gravitational instability \citep[GI][]{kratter16} and central binaries (with a stellar or brown dwarf secondary) have both been proposed as alternatives to planets in reproducing some of the observations --- \citep{dong15giarm} shows that GI can reproduce the observed symmetric pair of spiral arms \citep[see a similar example in the case of Elias 2-27;][]{meru17}\index[obj]{Elias 2-27}, and \citet{calcino20} used a central binary to explain the spirals as well as a few other features \citep[see also][]{price18, poblete19, poblete20}.

The most direct way to verify disk-planet interaction as the origin of observed disk structures is to find the planets predicted by the models. The best example here is PDS 70\index[obj]{PDS 70}, in which a wide gap $\sim60$ AU in radius has been found in scattered light and dust continuum observations \citep{hashimoto12, dong12pds70, hashimoto15}. \citet{zhu11,zhu12} showed that multiple planets may open a giant common gap much wider than the individual gap opened by a single planet. Together with dust filtration at the outer gap edge, such giant cavities, a.k.a. transitional disks, can be explained \citep[see also][]{duffell15dong, dong16td}. Two giant planets have since been found inside PDS 70's cavity \citep{keppler18, wagner18pds70,haffert19, hashimoto20}, and models show that they can well reproduce the observed disk structure \citep{bae19}. 

More recently, a planet was discovered around AB Aur \citep{2022NatAs...6..751C}\index[obj]{AB Aur}, with a projected separation of $\sim 93$ au. Interestingly, the disc around AB Aur contains spiral density waves observed with ALMA \citep{2017ApJ...840...32T} that may well be consistent with being caused by AB Aur b. It is possible that more planets exist closer to the star, that would help explain the dust cavity \citep{2017ApJ...840...32T} and some of the spiral features \citep{2020A&A...637L...5B}.

However, the success of PDS 70 and AB Aur are rare cases at the moment. In general, testing the planetary hypothesis for observed disk structures has been challenging. Planet candidates have been found in a few other systems and proposed to be the driver of observed disk structures, such as MWC 758 \citep{reggiani18, wagner19}, LkCa 15 \citep{sallum15}, and HD 100546 \citep{brittain14, currie15}. However their status remain uncertain at the moment. Direct imaging surveys have been carried out to systematically search for planets in disks \citep[e.g.,][]{asensiotorres21}. They remain largely inconclusive regarding the presence of the feature-producing planets \citep[e.g.,][]{dong18spiral}, partially due to the degeneracy in the planetary luminosity between hot and cold start models \citep{fortney08}.
The difficulty in finding these predicted giant planets may be expected, as they may experience episodic accretion and are only detectable in a small fraction of their formation lifetime \citep{brittain20}. The formation of some of these predicted planets, e.g., the multiple Saturn mass planets in the HL Tau disk at tens of AU and the supra-Jovian planet in the MWC 758 disk at over 100 AU, may also challenge our theoretical understanding on how planets form. The comparison between planet demographics at birth with those around billion years old stars obtained from radial velocity and transit surveys may also offer clues on the orbital evolution of planets.

Due to the difficulties in directly confirming the theory-predicted planets in disks, indirect approaches have been explored. For example, 
a brown dwarf or a stellar mass secondary can drive a pair of prominent arms in a symmetric configuration inside its orbit, in a way similar to a giant planet. In the case HD 100453, such an outer stellar companion (with a primary/secondary mass ratio of roughly 10:1) and a pair of spiral arms have been found \citep{chen06, wagner15hd100453, benisty17, vanderplas19}. The companion's orbit has been traced out \citep{wagner18}, and models and analysis have shown that the companion is likely responsible for driving the spirals \citep{dong16hd100453, wagner18, rosotti20}. The HD 100453 case is direct proof of an external companion exciting a pair of spirals as predicted by theory.

The pattern speed in asymmetric structures may also be used to test their disk-planet interaction origin. This has been attempted in the cases of MWC 758 and SAO 206462, both of which show a pair of symmetric arms \citep{benisty15, muto12, garufi13, stolker16sao206462}. Theory predicts that if the spirals are produced by a companion, the spirals and the companion should orbit the star at the same angular velocity (assuming the companion is on a circular orbit). \citet{ren18, ren20} measured the pattern speed of the two spirals in the MWC 758 disk, and concluded that they are consistent with being excited by an external companion at $\sim170$ AU. \citet{xie21} applied the same technique and measured the pattern speed in the SAO 206462 disk. They found that if the two spirals are co-moving, they are consistent with being excited by a planet on the outside at $\sim86$ AU. Meanwhile, current data does not provide strong constraints on whether they are co-moving or not, and if they are independently moving, they may be excited by two planets at $\sim120$ and 50 AU.
% SAO 206462: azimuthal shift.

\section{Summary and Conclusions}
\label{sec:con}

Despite having been around for more than 40 years \citep{1979ApJ...233..857G}, the theory of disk-planet interactions is still seeing significant progress. This is driven in part by an ever improving of our understanding of the intricate processes leading to angular momentum exchange between planet and disk (for example thermal and dynamical torques, and the impact of solids) and in part by a better understanding of protoplanetary disks, where a paradigm shift has occurred from viscous to wind-driven accretion disks. 

Since the last edition of Protostars and Planets, planet-disk interactions has become a subject that can be observed directly, rather than inferred from the current exoplanet population. We can use the theory of planet-disk interactions to interpret observations (see section \ref{sec:obs}), and a particular exciting prospect for the coming years is observation-driven improvement of the theory \citep[e.g.][]{nazari19}. This is necessary, since despite all recent progress, the problem of migration has not been 'solved'. If history is a guide, introducing new physics into the problem produces different torques that need to be understood. On the numerical side, given the small length scales involved in for example thermal torques, clever use of adaptive resolution may provide a way forward to understand how these small scales interact with the largest scales that in the end will determine how much mass can potentially flow towards the planet. The same holds for circumplanetary disks: the exciting observation of \cite{2021ApJ...916L...2B} means we need the most realistic CPD models possible, and probably resolve the planet at the same time \citep{2021arXiv210612003Z}. 

In view of all this, and the current influx of detailed observational data on protoplanetary disks and embedded planets, we expect again a lot of progress in our understanding of planet-disk interactions over the next few years.

%In addition to summarizing the main part of the chapter, this section will attempt to draw conclusions from the body of progress discussed, and to list the areas that are important to address, and/or are ripe for making progress, in the near future. This will be done in the wider context of all the areas covered by PPVII, attempting to show how future work in the planet-disk interactions field can advance wider science.

\vspace{5mm}
{\it Acknowledgements:} We thank Clement Baruteau, Sebastian Perez, Kazuhiro Kanagawa, Daniel Price, Farzana Meru, and Richard Nelson for helpful discussions. We thank Colin McNally for getting this chapter off the ground. We also want to thank Willy Kley (1958-2021), a true giant of disk-planet interactions. His contributions, both on a scientific and a personal level, have made an everlasting positive impact on the field.

\newpage
\bibliographystyle{pp7}
\bibliography{bibliography.bib}

%\printindex
%\renewcommand{\indexname}{Object Index}
%\printindex[obj]

\end{document}